\documentclass[prb,aps,superscriptaddress,twocolumn,longbibliography]{revtex4-2}
\usepackage{graphicx}
\usepackage{bm}
\usepackage{amsmath}
\usepackage{amsfonts}
\usepackage{amssymb}
\usepackage{epsfig}
\usepackage{color}
\usepackage{ulem}

\begin{document}

\title{Spin susceptibility for orbital-singlet Cooper pair \\
in the three-dimensional Sr$_2$RuO$_4$ superconductor}

%
\author{Yuri Fukaya}
\affiliation{Yukawa Institute for Theoretical Physics, Kyoto University, Kyoto 606-8502, Japan}
\affiliation{CNR-SPIN, I-84084 Fisciano (Salerno), Italy, c/o Universit\'a di Salerno, I-84084 Fisciano (Salerno), Italy}
\author{Tatsuki Hashimoto}
\affiliation{Yukawa Institute for Theoretical Physics, Kyoto University, Kyoto 606-8502, Japan}
\affiliation{Department of Mechanical Engineering, Stanford University, Stanford 94305, California, USA}
\author{Masatoshi Sato}
\affiliation{Yukawa Institute for Theoretical Physics, Kyoto University, Kyoto 606-8502, Japan}
\author{Yukio Tanaka}
\affiliation{Department of Applied Physics, Nagoya University, Nagoya 464-8603, Japan}
\author{Keiji Yada}
\affiliation{Department of Applied Physics, Nagoya University, Nagoya 464-8603, Japan}



\begin{abstract}
    We study the spin susceptibility of the orbital-singlet pairings, 
    including the spin-triplet/orbital-singlet/$s$-wave $E_g$ representation proposed 
    by Suh \textit{et al}., [H.\ G.\ Suh \textit{et al}., Phys.\ Rev.\ Research \textbf{2}, 032023 (2020)], 
    for a three-orbital model of superconducting Sr$_2$RuO$_4$ in three dimensions. 
    For the pseudospin-singlet states represented in the band basis, 
    the spin susceptibility decreases when reducing the temperature, 
    irrespective of the direction of the applied magnetic fields, 
    even if they are spin-triplet/orbital-singlet pairings in the spin-orbital space. 
    However, because the pseudospin-triplet \textbf{d}-vector in the band basis is not completely aligned in the $xy$-plane (along $z$-axis) owing to the strong atomic spin-orbit coupling,
    the spin susceptibility for spin-singlet/orbital-singlet/odd-parity pairings is reduced around $5$-$10$ percent with the decrease of the temperature along the $z$ ($x$) axis.
    We can determine the symmetry of the pseudospin structure of 
    the Cooper pair by the temperature dependence of the spin susceptibility measured 
    by nuclear magnetic resonance experiments. 
    Our obtained results serve as a guide to determine the pairing symmetry of Sr$_2$RuO$_4$. 
\end{abstract}

\maketitle

\section{Introduction}
Pairing symmetry in the Sr$_2$RuO$_4$ (SRO) superconductor (SC) 
~\cite{Maeno,MackenzieRMP2003,MaenoJPSJ2012}  
has been an unresolved issue in condensed matter physics.
Based on the previous various experiments, 
e.g., polarized neutron scatterings~\cite{DuffyPRL2000}, half-quantum vortices~\cite{JangScience2011,YasuiPRB2017},
charge transport properties in junctions
~\cite{YamashiroJPSJ,JinPRB1999,LaubePRL2000,Tanaka2009,WuPRB2010,KashiwayaPRL2011,AnwarNatPhys2016,OldePRB2018},
and the nuclear-magnetic-resonance (NMR) measurements~\cite{IshidaNature1998}, 
spin-triplet/chiral $p$-wave pairing with time-reversal symmetry (TRS) breaking
[($p_x+ ip_y$)-wave pairing]~\cite{RiceJPCM1995} has been believed to be 
the most promising one.
In addition, theoretical studies also supported the realization of 
spin-triplet/$p$-wave pairing~\cite{NomuraJPSJ2000,SatoJPSJ2000,TakimotoPRB2000,KurokiPRB2001,NomuraJPSJ2002,NomuraJPSJ2002_2,YanaseJPSJ2003,NomuraJPSJ2005,NomuraJPSJ2008,RaghuPRL2010,TsuchiizuPRB2015,ZhangPRB2018,WangPRL2019,WangPRB2020}.
However, recent NMR experiments, that solved the heating issues of the sample 
in the actual measurement process 
reported the reduction of the spin susceptibility 
with the in-plane magnetic field~\cite{PustogowNature2019,IshidaJPSJ2020,Chronistere2025313118}
below $T_\mathrm{c}$. 
These experiments seem to be inconsistent with spin-triplet/chiral $p$-wave 
where the \textbf{d}-vector is aligned along the $c$-axis
of SRO~\cite{Leggett2021}.


Experimental signatures of a two-component superconducting 
order parameter in SRO
were observed in ultrasound and thermodynamics experiments~\cite{GhoshNatPhys2021,AgterbergNatPhysics2021,BenhabibNatPhys2021}. 
Several theoretical studies focused on 
the two-component order parameter 
with TRS breaking:
the accidentally degenerate pairing 
[($s'+d_{x^2-y^2}$-wave~\cite{RomerPRL2019}, $d_{x^2-y^2}+ig_{xy(x^2-y^2)}$)-wave~\cite{KivelsonNPJ2020,WillaPRB2021,clepkens2021higher,YuanPRB2021}, and
($s+id_{xy}$)-wave~\cite{ClepkensPRR2021,RomerPRB2021}], 
and the interorbital $d_{zx}+id_{yz}$-like spin-triplet/orbital-singlet/$s$-wave $E_g$ pairing~\cite{SuhPRR2020}
with the Bogoliubov Fermi surface~\cite{AgterbergPRL2017,BrydonPRB2018}.
In the last case, 
the presence of the $t_{2g}$-orbital degrees of freedom and
strong atomic spin-orbit coupling in SRO~\cite{PuetterEPL2012,RamiresPRB2019,ChenPRB2020,SuhPRR2020}
can generate the orbital-singlet state. 
The pairing mechanism of the spin-triplet/orbital-singlet/$s$-wave pairing is 
due to the attractive channel $U'-J<0$ with the interorbital repulsive interaction $U'$ and 
the renormalized Hund's coupling $J$~\cite{PuetterEPL2012,SuhPRR2020}.
In addition, the recent experiment under hydrostatic pressure and disorder 
indicated the $d_{zx}+id_{yz}$-wave state~\cite{GrinenkoNatComm2021}.

To determine the spin structure of the Cooper pair,
the temperature dependence of the spin susceptibility 
in the NMR experiments gives us the important information 
~\cite{IshidaNature1998,PustogowNature2019,IshidaJPSJ2020,Chronistere2025313118}.
In the theoretical approaches in SRO,
spin susceptibility was calculated 
in the spin-triplet/orbital-singlet/$s$-wave 
pairing with the \textbf{d}-vector along the $z$-axis 
as a function of the temperature~\cite{YuPRB2018} 
and under the uniaxial strain~\cite{LindquistPRR2020}
in the two-dimensional multiorbital SRO model.
Since there are three $t_{2g}$-orbitals near the Fermi level in SRO, 
it is necessary to study the temperature dependence of the spin susceptibility 
for the possible orbital-singlet Cooper pair taking into account the 
orbital nature.
Then, we must adopt the ``three-dimensional'' SRO Hamiltonian
to investigate 
the orbital-singlet $d_{zx}+ id_{yz}$-like pair potential~\cite{SuhPRR2020}.

In this paper, we calculate the temperature dependence of the spin susceptibility
below the critical temperature $T_\mathrm{c}$ for orbital-singlet pairings 
in the three-dimensional SRO model by choosing the possible irreducible 
representations. 
We focus on the spin-triplet/orbital-singlet/$s$-wave
and spin-singlet/orbital-singlet/odd-parity pairings 
stemming from the multiorbital
and strong atomic spin-orbit coupling. 
In the first case, the pseudospin-singlet pairing is realized 
in the band basis, then the resulting 
spin susceptibility is reduced with the decrease of temperature 
irrespective of the direction of the magnetic field for all possible irreducible representations. 
In the second case, 
the spin susceptibility changes around $5\%$ ($10\%$) by the temperature along the $x$ ($z$) axis, because the pseudospin-triplet \textbf{d}-vector in the band basis is not perfectly aligned in the $xy$-plane ($z$-axis) away from the $xy$-symmetric plane owing to the strong atomic spin-orbit coupling. 
We conclude that the recently observed spin susceptibility of NMR experiments
in SRO~\cite{PustogowNature2019,IshidaJPSJ2020,Chronistere2025313118}
can be explained by the spin-triplet/orbital-singlet/$s$-wave $E_g$ representation. 

\section{Model Hamiltonian and formulation}
In this section, we show the model Hamiltonian and the
formulation to calculate the spin susceptibility in SRO. 

\begin{table*}[htbp]
    \caption{Classification of the orbital-singlet pairings in the point group $D_{4h}$~\cite{SuhPRR2020}. 
    Spin-triplet/orbital-singlet/even-parity (TSE) pairing $\hat{\Delta}=\Delta(T)[\hat{L}_{i}\otimes\hat{\sigma}_{j}]i\hat{\sigma}_{y}$ 
    is described by the \textbf{d}-vector.
    Spin-singlet/orbital-singlet/odd-parity (SSO) pairing is expressed by the spin-singlet pair potential 
    $\hat{\Delta}(\bm{k})=\Delta(T)[\hat{L}_{i}\otimes\hat{\sigma}_{0}\sin{k}_{j}]i\hat{\sigma}_{y}$ for $[i,j]$ $(i,j=x,y,z)$. 
    Here $[i,j]$ means the indices of $\hat{L}_{i}$ and $\hat{\sigma}_{j}$ in TSE pairing, and $\hat{L}_{i}$ and ${k}_{j}$ in SSO, respectively.
    We focus on the even-frequency pair potential in this table. }
    \label{table_1}
    \begin{center}
        \begin{tabular}{cccc}
            Irreducible rep. & State & Pair potential $[i,j]$ & Gap structure \\ \hline
            $A_{1g}$ & TSE & $[y,y]+[x,x]$ & Fully gapped \\
            $A_{1g}$ & TSE & $[z,z]$ & fully Gapped \\
            $A_{2g}$ & TSE & $[y,x]-[x,y]$ & Gapless \\
            $B_{1g}$ & TSE & $[y,y]-[x,x]$ & Line node in the diagonal directions \\
            $B_{2g}$ & TSE & $[y,x]+[x,y]$ & Line node in the $x$ and $y$ directions \\
            $E_{g}$ & TSE & $\{[z,x],[z,y]\}$ & Bogoliubov Fermi surface in $k_z=0,2\pi$ planes \\ 
            $E_{g}$ & TSE & $\{[x,z],[y,z]\}$ & Bogoliubov Fermi surface in $k_z=0,2\pi$ planes~\cite{SuhPRR2020} \\ \hline
            $A_{1u}$ & SSO & $[y,y]+[x,x]$ & Fully gapped \\
            $A_{1u}$ & SSO & $[z,z]$ & Line node in $k_z=0,2\pi$ planes \\
            $A_{2u}$ & SSO & $[y,x]-[x,y]$ & Line node in the $x$ and $y$ directions \\
            $B_{1u}$ & SSO & $[y,y]-[x,x]$ & Fully gapped \\
            $B_{2u}$ & SSO & $[y,x]+[x,y]$ & Line node in the $x$ and $y$ directions \\ 
            $E_{u}$ & SSO & $\{[z,x],[z,y]\}$ & Bogoliubov Fermi surface in $zx$ and $yz$ planes \\
            $E_{u}$ & SSO & $\{[x,z],[y,z]\}$ & Bogoliubov Fermi surface in $k_z=0,2\pi$ planes \\ \hline
        \end{tabular}
    \end{center}
\end{table*}%
SRO has the $I4/mmm$ tetragonal space group
with the point group $D_{4h}$~\cite{MackenzieRMP2003}.
The conduction bands of SRO mainly consist of
$t_{2g}$-orbitals [$d_{yz}$, $d_{zx}$, and $d_{xy}$] in the Ru ions.
The Hamiltonian in SRO is written as 
\begin{align}
    \hat{\mathcal{H}}=\sum_{\bm{k}}\hat{C}^{\dagger}_{\bm{k}}\hat{H}(\bm{k})\hat{C}_{\bm{k}},
    \label{H1}
\end{align}%
where $\hat{C}^{\dagger}_{\bm{k}}=[c^{\dagger}_{yz,\uparrow\bm{k}},c^{\dagger}_{zx,\uparrow\bm{k}},c^{\dagger}_{xy,\uparrow\bm{k}},c^{\dagger}_{yz,\downarrow\bm{k}},c^{\dagger}_{zx,\downarrow\bm{k}},c^{\dagger}_{xy,\downarrow\bm{k}}]$
is the creation operator of electrons in $t_{2g}$-orbitals.
For $\hat{\mathcal{H}}$ in Eq.\ (\ref{H1}),
we adopt the three-dimensional Hamiltonian in Refs.~\cite{ScaffidiPRB2014,RamiresPRB2016,HaverkortPRL2008,VeenstraPRL2014,RamiresPRB2019,RosisingPRR2019,SuhPRR2020,ClepkensPRR2021},
\begin{align}
    \hat{H}(\bm{k})&=\sum_{l,j}h_{lj}(\bm{k})\hat{\Lambda}_{l}\otimes\hat{\sigma}_{j},
\end{align}%
where $\hat{\Lambda}_{l=0\sim 8}$ are the Gell-Mann matrices as shown in Appendix A,
and $\hat{\sigma}_{j=0,x,y,z}$ are the Pauli ones in the spin space.
[The explicit form of $h_{lj}(\bm{k})$ is given in Appendix A.] 

In the superconducting state, the Bogoliubov$-$de Gennes (BdG) Hamiltonian is given by
\begin{align}
    \hat{H}_\mathrm{BdG}(\bm{k})&=
    \begin{pmatrix}
        \hat{H}(\bm{k}) & \hat{\Delta}(\bm{k}) \\
        \hat{\Delta}^{\dagger}(\bm{k}) & -\hat{H}^{*}(-\bm{k})
    \end{pmatrix},
\end{align}%
with the pair potential (energy gap function) $\hat{\Delta}(\bm{k})$. 
Here, we consider the pair potential by the symmetry of the Cooper pair.
The present model Hamiltonian has the parity dependence in $\bm{k}$
and spin-orbital degrees of freedom. 
Then the pair potential can classify the four types of Cooper pair
that satisfy the Fermi-Dirac statistics:
spin-singlet/orbital-triplet/even-parity (STE), 
spin-triplet/orbital-triplet/odd-parity (TTO), 
spin-triplet/orbital-singlet/even-parity (TSE),
and spin-singlet/orbital-singlet/odd-parity (SSO). 
In our study, we focus on the orbital-singlet pair potentials, i.e., TSE and SSO. 
Note that we do not consider the odd-frequency pairing in the pair potential
because we do not adopt the retardation effect in the attractive channel
~\cite{Berezinskii,Balatsky,Shigeta,tanaka12,LinderRMP2019}.
For TSE states, we assume the ``\textit{isotropic}'' pairing and
the energy gap function is independent of $\bm{k}$.
The TSE states are described by the spin-triplet potentials~\cite{PuetterEPL2012,RamiresPRB2019,RosisingPRR2019,SuhPRR2020},
\begin{align}
    \hat{\Delta}&=\Delta(T)[\hat{L}_{i}\otimes\hat{\sigma}_{j}]i\hat{\sigma}_{y},
    \label{TSE}
\end{align}%
with $i,j=x,y,z$ and
the $t_{2g}$-orbital anglar momentum operators projected onto $L=2$
in the $[d_{yz},d_{zx},d_{xy}]$ basis,
\begin{align*}
    \hat{L}_{x}&=
    \begin{pmatrix}
        0 & 0 & 0 \\
        0 & 0 & i \\
        0 & -i & 0
    \end{pmatrix}, \hspace{1mm}
    \hat{L}_{y}=
    \begin{pmatrix}
        0 & 0 & -i \\
        0 & 0 & 0 \\
        i & 0 & 0
    \end{pmatrix}, \hspace{1mm}
    \hat{L}_{z}=
    \begin{pmatrix}
        0 & i & 0 \\
        -i & 0 & 0 \\
        0 & 0 & 0
    \end{pmatrix},
\end{align*}%
respectively.
Here, we define the indices of $\hat{L}_{i}$ and $\hat{\sigma}_{j}$ in Eq.\ (\ref{TSE}) as $[i,j]$.
$\Delta(T)$ is the pair potential at the temperature $T$ and
it has the Bardeen-Cooper-Schrieffer (BCS)-like temperature dependence,
\begin{align}
    \Delta(T)&=\alpha_{c}\Delta_{0}\tanh\left[1.74\sqrt{\frac{T_\mathrm{c}-T}{T}}\right],\label{T_dep_gap}\\
   \Delta_{0}&=\frac{3.53}{2}T_\mathrm{c},
\end{align}%
with the critical temperature $T_\mathrm{c}$.
We choose $\alpha_{c}$ so that
the maximal quasiparticle energy gap amplitude
becomes $\Delta_0$,
and its value is given in Appendix C (Table II).
Likewise, for SSO pairings, we consider the spin-singlet pair potentials,
\begin{align}
    \hat{\Delta}(\bm{k})&=\Delta(T)[\hat{L}_{i}\otimes\hat{\sigma}_{0}\sin{k}_{j=x,y}a]i\hat{\sigma}_{y}, \\
    \hat{\Delta}(\bm{k})&=\Delta(T)\left[\hat{L}_{i}\otimes\hat{\sigma}_{0}\sin\frac{{k}_{j=z}c}{2}\right]i\hat{\sigma}_{y},
\end{align}%
with the lattice constants $[a,a,c]$ and the definition of the indices $\hat{L}_{i}$ and ${k}_{j=x,y,z}$ as $[i,j]$.
Table~\ref{table_1} shows the classification of
orbital-singlet pair potentials.
We obtain 14 orbital-singlet pair potentials
for both TSE and SSO states in the point group $D_{4h}$. 
Only interorbital $E_g$ and $E_u$ representations can break the TRS
among the orbital-singlet pairings in Table~\ref{table_1}.
The TRS broken pairings for TSE $E_g$ $\{[z,x],[z,y]\}$ and $\{[x,z],[y,z]\}$ representations
are written by the linear combination,
\begin{align}
    \hat{\Delta}
    &=\Delta(T)[\hat{L}_{z}\otimes(\hat{\sigma}_{x}+i\hat{\sigma}_{y})]i\hat{\sigma}_{y}, \\
    \hat{\Delta}
    &=\Delta(T)[(\hat{L}_{x}+i\hat{L}_{y})\otimes\hat{\sigma}_{z}]i\hat{\sigma}_{y}, \label{gap_Eg}
\end{align}%
respectively.
Likewise, the time-reversal broken pairings for SSO $E_u$ $\{[z,x],[z,y]\}$
and $\{[x,z],[y,z]\}$ representations are given by
\begin{align}
    \hat{\Delta}(\bm{k})
    &=\Delta(T)[\hat{L}_{z}\otimes\hat{\sigma}_{0}(\sin{k_{x}a}+i\sin{k_{y}a})]i\hat{\sigma}_{y}, \\
    \hat{\Delta}(\bm{k})
    &=\Delta(T)\left[(\hat{L}_{x}+i\hat{L}_{y})\otimes\hat{\sigma}_{0}\sin\frac{{k_{z}c}}{2}\right]i\hat{\sigma}_{y}. 
\end{align}%

Spin susceptibility $\chi_{i}(T)$ along the $i=x,y,z$ axis at temperature $T$
is given by the Kubo formula~\cite{HirashimaJPSJ2007,Maruyama2012JPSJ,HashimotoJPSJ2013},
\begin{align}
    \chi_{i}(T)&=T\int_{\mathrm{BZ}}d\bm{k}\sum_{i\varepsilon_{n}}
    \mathrm{Tr}[\hat{s}_{i}\hat{g}(\bm{k},i\varepsilon_n)\hat{s}_{i}\hat{g}(\bm{k},i\varepsilon_n)]\\
    &=\int_\mathrm{BZ}d\bm{k}\sum_{\alpha,\beta}\langle\alpha|\hat{s}_{i}|\beta\rangle
    \langle\beta|\hat{s}_{i}|\alpha\rangle \notag \\
    &\times T\sum_{i\varepsilon_n}G_{\alpha}(\bm{k},i\varepsilon_n)G_{\beta}(\bm{k},i\varepsilon_n)
    e^{+i\varepsilon_{n}0}, 
\end{align}%
\begin{align}
    \hat{g}(\bm{k},i\varepsilon_n)&=\frac{1}{i\varepsilon_n-\hat{H}_\mathrm{BdG}(\bm{k})},\\
    \hat{H}_\mathrm{BdG}(\bm{k})|\alpha\rangle&=E_{\alpha}(\bm{k})|\alpha\rangle,
\end{align}%
where $\hat{s}_{i=x,y,z}$ are the spin angular momentum operators 
expanded in particle-hole space, 
$i\varepsilon_n=i(2n+1)\pi T$ is the fermionic Matsubara frequency, 
$E_{\alpha(\beta)}(\bm{k})$ is the Bogoliubov energy band, and
$|\alpha(\beta)\rangle$ is the eigenstate corresponding to the 
Bogoliubov energy band $E_{\alpha(\beta)}(\bm{k})$ with the band indices $\alpha,\beta$. 
Here, $\hat{g}(\bm{k},i\varepsilon_n)$ stands for the matrix of the Green's function
in the spin-orbital basis and
$G_{\alpha}(\bm{k},i\varepsilon_n)$ denotes the Green's function
defined by
\begin{align}
    G_{\alpha}(\bm{k},i\varepsilon_n)&=\frac{1}{i\varepsilon_n-E_{\alpha}(\bm{k})}.
\end{align}%
Here, we adopt the formulation,
\begin{align}
    &T\sum_{i\varepsilon_n}G_{\alpha}(\bm{k},i\varepsilon_n)G_{\beta}(\bm{k},i\varepsilon_n)
    e^{+i\varepsilon_{n}0} \notag \\
    &=
    \begin{cases}
        -\frac{1}{4T}
        \left[1-\tanh^{2}\frac{E_{\alpha}(\bm{k})}{2T}\right] & E_{\alpha}(\bm{k})=E_{\beta}(\bm{k})\\
        -\frac{\tanh\frac{E_{\alpha}(\bm{k})}{2T}-\tanh\frac{E_{\beta}(\bm{k})}{2T}}
        {2[E_{\alpha}(\bm{k})-E_{\beta}(\bm{k})]} & E_{\alpha}(\bm{k})\neq E_{\beta}(\bm{k})
    \end{cases},
\end{align}%
to sum up the Matsubara frequency from $-\infty$ to $\infty$ analytically.
Although the Fermi surface along the $z$-axis is almost cylindrical~\cite{SuhPRR2020}
and the $t_{2g}$-orbital characters at the Fermi level
are nearly independent of $k_z$ [see also Appendix B (Fig.~\ref{AFig1})], 
we need the integration of $k_z$ for all representations in the actual calculation. 

\begin{figure*}[htbp]
    \centering
    \includegraphics[width=17.5cm]{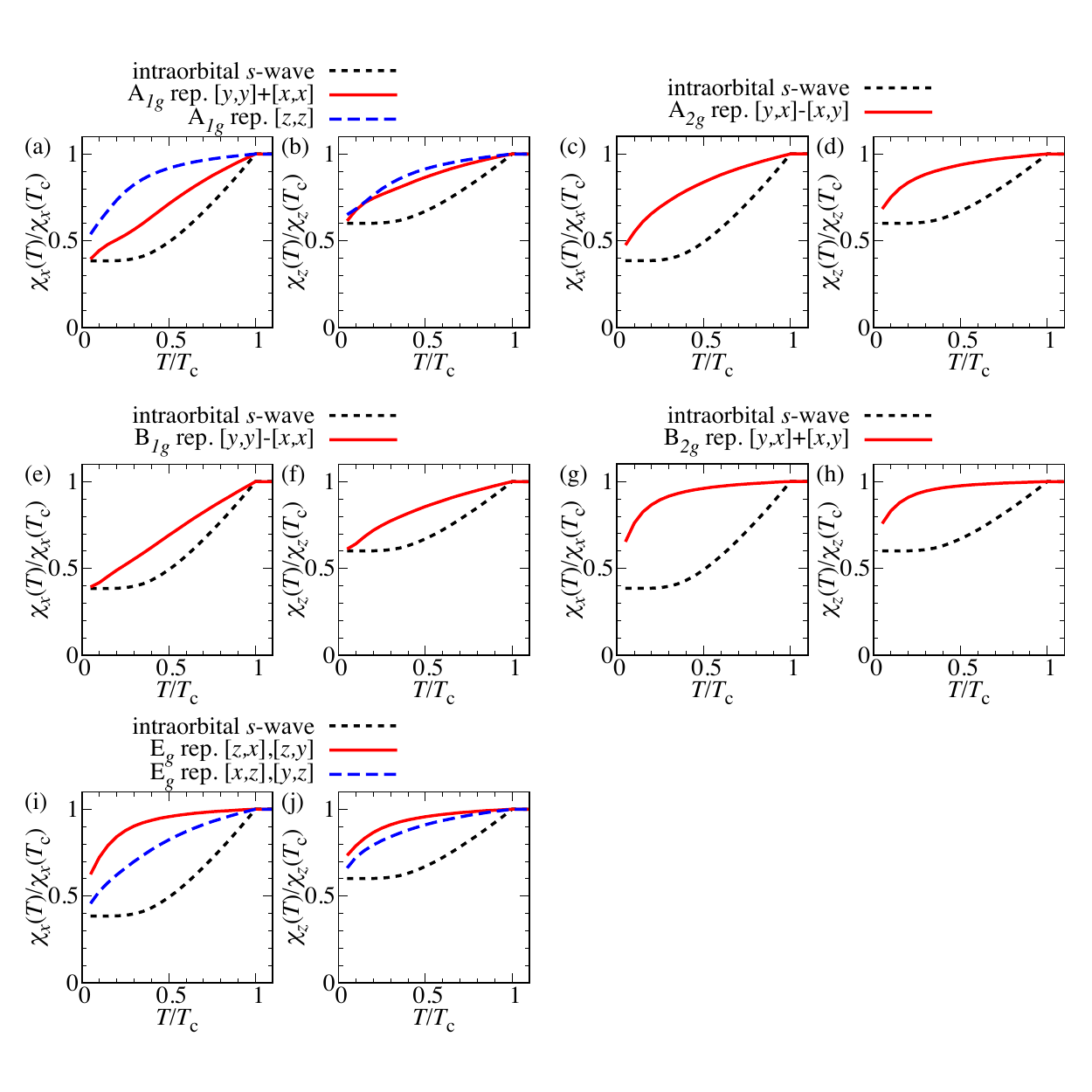}
    \caption{Spin susceptibility $\chi_{i=x,z}(T)$ for intraorbital spin-singlet $s$-wave (black dotted line) and spin-triplet/orbital-singlet/$s$-wave (TSE) pairings normalized by $\chi_{i}(T_\mathrm{c})$
    along the (a), (c), (e), (g), and (i) $x$, and (b), (d), (f), (h), and (j) $z$-directions
    as a function of the temperature.
    As shown in Table.~\ref{table_1}, we choose the pair potential as
    (a,b) TSE $A_{1g}$ $[y,y]+[x,x]$ (red solid line) and $[z,z]$ (blue dotted line),
    (c,d)$A_{2g}$, (e,f)$B_{1g}$, (g,h) $B_{2g}$, 
    and $E_{g}$ $\{[z,x],[z,y]\}$ (red solid line) and $\{[x,z],[y,z]\}$ (blue dotted line) states, respectively.
    Here, we do not plot $\chi_{y}(T)/\chi_{y}(T_\mathrm{c})$
    because spin susceptibility along the $y$ direction $\chi_{y}(T)$ is the same as
    that along the $x$ axis $\chi_{x}(T)$
    in the presence of the fourfold rotational symmetry in the $xy$ plane. 
    }
    \label{fig1}
\end{figure*}%
\begin{figure*}[htbp]
    \centering
    \includegraphics[width=17.5cm]{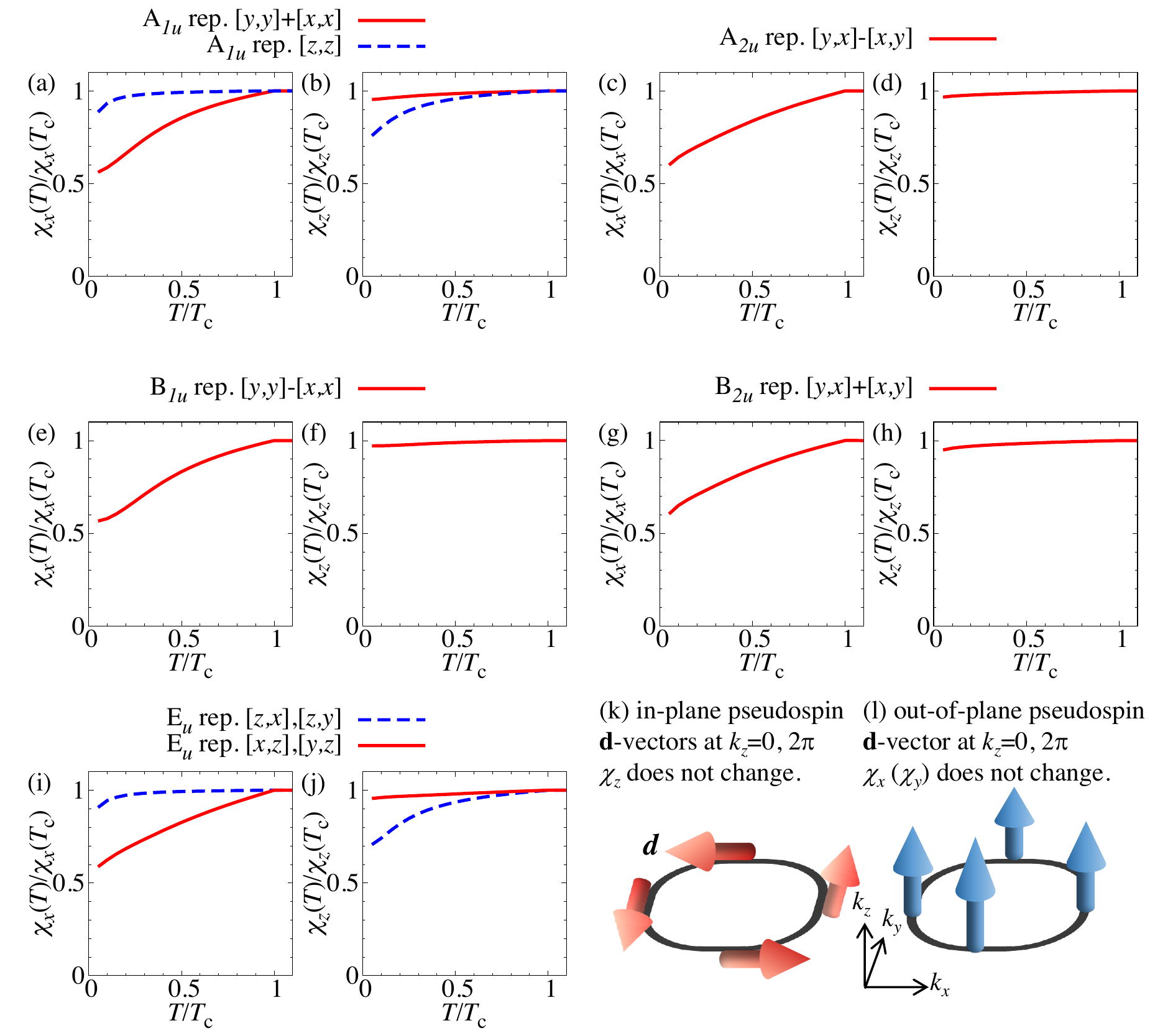}
    \caption{Spin susceptibility $\chi_{i=x,z}(T)$ for spin-singlet/orbital-singlet/odd-parity (SSO) pairings normalized by $\chi_{i}(T_\mathrm{c})$
    along the (a), (c), (e), (g), and (i) $x$, and (b), (d), (f), (h), and(j) $z$ axes
    as a function of the temperature.
    As shown in Table.~\ref{table_1}, we select the pair potential as
    (a,b) SSO $A_{1u}$ $[y,y]+[x,x]$ (red solid line) and $[z,z]$ (blue dotted line),
    (c,d)$A_{2u}$, (e,f)$B_{1g}$, (g,h) $B_{2u}$, 
    and (i,j) $E_{u}$ $\{[z,x],[z,y]\}$ (blue dotted line) and $\{[x,z],[y,z]\}$ (red solid) states, respectively.
    Schematic illustration of (k) in-plane and (l) out-of-plane pseudospin \textbf{d}-vectors in the band basis at $k_z=0,2\pi$.
    Black line means the Fermi line.
    Away from the $xy$-symmetric plane, pseudospin \textbf{d}-vector is not aligned in the $xy$plane and along the $z$ axis, respectively.
    }
    \label{fig2}
\end{figure*}%

\section{Results and discussion}
We show the temperature dependence of the calculated spin susceptibility
below the critical temperature $T_\mathrm{c}$
for the orbital-singlet pairings in the three-dimensional SRO model.
Figure~\ref{fig1} shows the temperature dependence of the spin susceptibility $\chi_{i=x,z}(T)$
normalized by $\chi_{i}(T_\mathrm{c})$ where the direction of the applied field
is along the $x$-axis for Figs.~\ref{fig1}(a), \ref{fig1}(c), \ref{fig1}(e), \ref{fig1}(g), and \ref{fig1}(i), and the $z$-axis for Figs.~\ref{fig1}(b), \ref{fig1}(d), \ref{fig1}(f), \ref{fig1}(h), and \ref{fig1}(j).
The spin susceptibility along the $y$-direction is the same as
that along the $x$-axis due to the fourfold rotational symmetry in the $xy$-plane.
In Fig.~\ref{fig1}, the pair potentials used in the calculation are interorbital TSE
$A_{1g}$ [Figs.~\ref{fig1}(a) and \ref{fig1}(b)], $A_{2g}$ [Figs.~\ref{fig1}(c) and \ref{fig1}(d)], $B_{1g}$ [Figs.~\ref{fig1}(e) and \ref{fig1}(f)], $B_{2g}$ [Figs.~\ref{fig1}(g) and \ref{fig1}(h)], and $E_{g}$ [Figs.~\ref{fig1}(i) and \ref{fig1}(j)] representations. 
The calculation result for intraorbital spin-singlet $s$-wave state (BCS state)
is also shown for reference in Fig.~\ref{fig1} (black dotted line).
Note that TSE $E_g$ $\{[x,z],[y,z]\}$ representation
with TRS breaking
is one of the promising candidates of pairing symmetry in SRO
and the resulting energy spectrum has the Bogoliubov Fermi surface in $xy$-plane~\cite{SuhPRR2020}.
It is also noted that nonzero atomic spin-orbit coupling needs to open the energy gap.
For the TSE state in Fig.~\ref{fig1},
spin susceptibility decreases as temperature decreases
for any irreducible representation in Table.~\ref{table_1}
for both the $x$- and $z$-directed applied magnetic fields.
In addition, as the quasiparticle energy spectrum in interorbital pairings does not open the energy gap $\Delta(T)$ on the Fermi surface,
even if $\Delta(T)$ is modified by $\alpha_c$ in the BdG Hamiltonian,
the function of the spin susceptibility $\chi_{i}(T)$ for interorbital pairings is convex upwards, not downwards.

On the other hand, for the interorbital SSO pairings as shown in Fig.~\ref{fig2},
the temperature dependence of the spin susceptibility
is sensitive to the direction of the applied magnetic field.
In the case of $A_{1u}$ $[y,y]+[x,x]$ [red solid line in Figs.~\ref{fig2}(a) and \ref{fig2}(b)], $A_{2u}$ [Figs.~\ref{fig2}(c) and \ref{fig2}(d)], $B_{1u}$ [Figs.~\ref{fig2}(e) and \ref{fig2}(f)], $B_{2u}$[Figs.~\ref{fig2}(g) and \ref{fig2}(h)], and $E_u$ $\{[x,z],[y,z]\}$ [red solid line in Figs.~\ref{fig2}(i) and \ref{fig2}(j)],
the spin susceptibility decreases $\sim 50\%$ when the direction of the field is in the in-plane, and $\sim 5\%$ along $z$-axis, as shown in Fig.~\ref{fig2}.
In contrast, spin susceptibility for SSO $A_{1u}$ $[z,z]$ [blue dotted line in Figs.~\ref{fig2}(a) and \ref{fig2}(b)] and $E_u$ $\{[z,x],[z,y]\}$ [blue dotted line in Figs.~\ref{fig2}(i) and \ref{fig2}(j)] representations
decreases for the magnetic field in the $z$-direction, as well as $\sim 10\%$ along the $x$-axis at low temperature.
These results contradict naive understanding for the single-band results
where spin susceptibility decreases for any direction of the field 
in a spin-singlet pairing or along the direction parallel to the \textbf{d}-vector of the pair potential.

To resolve this,
we focus on the pseudospin state in the band basis and parity dependence
for each orbital-singlet pair potential.
In principle, when the spin-singlet/even-parity pairing is realized in the single orbital model,
spin susceptibility goes to zero at $T=0$ irrespective of the direction of the magnetic field.
In the spin-triplet/odd-parity pairing, 
the spin susceptibility is reduced below $T_\mathrm{c}$
if the \textbf{d}-vector is parallel to the magnetic field.
However, when the \textbf{d}-vector is perpendicular to the magnetic field,
spin susceptibility does not change with the temperature.
It implies that we can determine the spin structure by the temperature dependence of the spin susceptibility in the single-orbital model.
In the present study, spin susceptibility for TSE (SSO) pairings
is reduced independently of the direction of the applied magnetic fields
(has the anisotropic behavior for the directions).
For TSE pairings, the temperature dependence
shown in Fig.~\ref{fig1}
is caused by the pseudospin-singlet state in the band basis,
despite the spin-triplet pairing in the spin-orbital space.
We note that spin susceptibility does not go to zero at $T=0$
due to the Van-Vleck paramagnetism
in the presence of the atomic spin-orbit coupling.
In SSO pairings as shown in Fig.~\ref{fig2}, we can adopt the \textbf{d}-vector
that describes the pseudospin-triplet state.
As shown in Fig.~\ref{fig2}, the spin susceptibility decreases $5$-$10\%$ along the direction where there is no reduction in the single-orbital model.
In the present study, this pseudospin \textbf{d}-vector is not perfectly aligned in the $xy$-plane or along the $z$-axis away from the $xy$-symmetric plane owing to the strong atomic spin-orbit coupling.
Since the pseudospin \textbf{d}-vector is almost in-plane
in the SSO $A_{1u}$ $[y,y]+[x,x]$, $A_{2u}$, $B_{1u}$, $B_{2u}$, and $E_u$ $\{[x,z],[y,z]\}$ representations,
the spin susceptibility changes around $5\%$ along the $z$-axis.
For the $A_{1u}$ $[z,z]$ and $E_u$ $\{[z,x],[z,y]\}$ representations,
the pseudospin \textbf{d}-vector is out-of-plane and it is not parallel to the $z$-axis. 
Thus, spin susceptibility in the $A_{1u}$ $[z,z]$ and $E_u$ $\{[z,x],[z,y]\}$ pairings is reduced $\sim 10\%$ by in-plane applied magnetic field at low temperature.
These behaviors also occur even in the intraorbital spin-triplet/odd-parity and the interorbital TTO pairings. 
We show the spin susceptibility for the intraorbital chiral $p$-wave pairing in Appendix F, on behalf of all spin-triplet/odd-parity states.
The spin susceptibility along the $z$-direction does not become zero owing to the strong atomic spin-orbit coupling in the interorbital $A_{1u}$ $[z,z]$ and $E_u$ $\{[z,x],[z,y]\}$ states. 

Here, we point out the relation between spin susceptibility and pseudospin/parity state.
The behavior with the temperature
in TSE (SSO) pairings is similar to that in spin-singlet/even-parity
(spin-triplet/odd-parity).
We can mention that the symmetry of the parity coincides with
the temperature dependence of the spin susceptibility for the orbital-singlet pair potential. 
Therefore, in multiorbital SCs with strong atomic spin-orbit coupling, 
the temperature dependence of the spin susceptibility 
for the orbital-singlet Cooper pair
is determined by the pseudospin/parity state in the band basis.
For this perspective, we can mention that
spin susceptibility for orbital-singlet pairings with different momentum dependence, e.g., TSE $d$ and SSO $f$-wave, behaves qualitatively the same as that for $s$ and $p$-wave cases in the present study, respectively.
We note that these kinds of temperature dependence for orbital-singlet pairings in the present study
are the same as a theoretical research of the spin susceptibility
in the superconducting topological insulator Cu$_x$Bi$_2$Se$_3$~\cite{HashimotoJPSJ2013}.

\section{Summary and conclusion}
We studied the temperature dependence of the spin susceptibility 
below $T_\mathrm{c}$
for the orbital-singlet Cooper pair in SRO. 
The pseudospin state in the band basis is determined by the parity of the pair potential.
In other words, the pseudospin-singlet (triplet) state is
realized in the case of even (odd) parity pairing.
If we consider orbital-singlet pairing, 
pseudospin-singlet (triplet) state means spin-triplet (singlet) pairing.
Thus, the spin susceptibility for the spin-triplet/orbital-singlet/$s$-wave pairings
decreases with the temperature,
independently of the direction of the applied magnetic fields.
In the spin-singlet/orbital-singlet/odd-parity pairings,
the spin susceptibility decreases around $5\%$ ($10\%$) along the $z$ ($x$) axis
for $A_{1u}$ $[y,y]+[x,x]$, $A_{2u}$, $B_{1u}$, $B_{2u}$, and $E_{u}$ $\{[x,z],[y,z]\}$ ($A_{1u}$ $[z,z]$, and $E_{u}$ $\{[z,x],[z,y]\}$) representations at low temperature.
It is caused by the pseudospin \textbf{d}-vector that is not completely aligned in the $xy$-plane (along the $z$-direction) away from the $xy$-symmetric plane due to the strong atomic spin-orbit coupling.
This behavior is relevant to the effect of the atomic spin-orbit coupling, not the orbital nature in the superconducting state. 
Here, the quantitative of the spin susceptibility strongly depends on the
length of the atomic spin-orbit coupling, 
as the importance of the strong spin-orbit coupling in SRO was pointed out~\cite{HaverkortPRL2008,VeenstraPRL2014}.
Based of the present study, 
the recent NMR experiments~\cite{PustogowNature2019,IshidaJPSJ2020,Chronistere2025313118} 
indicate not the spin-singlet pairing, but the pseudospin-singlet/even-parity one in SRO.
At least, since the spin-triplet/orbital-singlet/$s$-wave pairings are pseudospin-singlet states, they do not contradict the recent NMR experiments~\cite{PustogowNature2019,IshidaJPSJ2020,Chronistere2025313118}. 
Likewise, the spin susceptibility for accidentally degenerate intraorbital spin-singlet pairings~\cite{RomerPRL2019,KivelsonNPJ2020,WillaPRB2021,clepkens2021higher,YuanPRB2021,ClepkensPRR2021,RomerPRB2021}, that behaves the same as that for the intraorbital spin-singlet cases, is also consistent with these NMR experiments. 
To elucidate the pairing symmetry of the spin-degree of freedom 
of the present spin-triplet/orbital-singlet/even-parity and spin-singlet/orbital-singlet/odd-parity pairings, 
charge transport in SC/ferromagnet junctions with a well-oriented interface 
is highly desired~\cite{KTYB99,Hirai03}
because tunneling spectroscopy via Andreev bound states plays an important role in the 
determination of the unconventional superconductors~\cite{TK95,KashiwayaTanaka2000RepProgPhys}.

\section{Acknowledgements}
This work is supported by the JSPS KAKENHI (Grants No.\ JP15H05851, No.\ JP15H05853, 
No.\ JP15K21717, No.\ JP18H01176, No.\ JP18K03538, No.\ JP20H00131, and
No.\ JP20H01857) from MEXT of Japan, 
Researcher Exchange Program between JSPS and RFBR (Grants No.\ JPJSBP120194816), 
and the JSPS Core-to-Core program Oxide Superspin international network (Grants No.\ JPJSCCA20170002).
We thank Y.\ Maeno, P.\ Gentile, and H.\ Kaneyasu for the helpful comments.
We also appreciate the valuable comments and discussions by H.\ G.\ Suh and D.\ F.\ Agterberg.

\appendix

\section{Model Hamiltonian of three-dimensional Sr$_2$RuO$_4$ in the normal state}

\begin{table*}[htbp]
    \caption{Parameters in three-dimensional Sr$_2$RuO$_4$ model in Ref.~\cite{SuhPRR2020}.
    We set all values in meV.}
    \label{table_A1}
    \begin{center}
        \begin{tabular}{ccccccccccccc} \hline
            $t^{(z,z)}_{x}=-362.4$ & & $t^{(z,z)}_{y}=-134$ & & $t^{(xy,xy)}_{x}=-262.4$ & & $t^{(z,z)}_{xy}=-44.01$ & &
            $t^{(z,z)}_{xx}=-1.021$ & & $t^{(z,z)}_{yy}=-5.727$ & & $t^{(xy,xy)}_{xy}=-43.73$ \\
            $t^{(xy,xy)}_{xx}=34.23$ & & $t^{z}_{xy}=16.25$ & & $t^{(z,z)}_{xxy}=-13.93$ & & $t^{(z,z)}_{xyy}=-7.52$ & & $t^{(xy,xy)}_{xxy}=8.069$ & &
            $t^{z}_{xxy}=3.94$ & & $\lambda_\mathrm{SO}=57.39$  \\
            $\mu_{z}=438.5$ & & $\mu_{xy}=218.6$ & & $t^{(z,z)}_{z}=-0.0228$ & & $t^{(xy,xy)}_{z}=1.811$ & & $t^{z}_{z}=9.975$ & & $t^{(zx,xy)}_{z}=8.304$ & &
            $t^{(z,z)}_{zz}=2.522$ \\
            $t^{(xy,xy)}_{zz}=-3.159$ & & $\lambda^\mathrm{SOC}_{56z}=-1.247$ & & $\lambda^\mathrm{SOC}_{12z}=-3.576$ & &
            $\lambda^\mathrm{SOC}_{5162}=-1.008$ & & $\lambda^\mathrm{SOC}_{5261}=0.3779$ & & & & \\ \hline
        \end{tabular}
    \end{center}
\end{table*}
\begin{table*}[htbp]
    \caption{$\Delta_\mathrm{eff}/\Delta_{0}$ for each energy band.
    We choose the maximum value of $\Delta_\mathrm{eff}/\Delta_{0}$
    as $\alpha_c=\Delta_0/\Delta_\mathrm{eff}$.
    For the interorbital A$_{2g}$ representation, the gapless state appears.}
    \label{table_A2}
    \begin{center}
        \begin{tabular}{cccccc}
            Irreducible rep. & State & Gap function & $\alpha$-band & $\gamma$-band & $\beta$-band \\ \hline
            $A_{1g}$ & TSE & $[y,y]+[x,x]$ & 0.167 & 1.01 & 1.27 \\
            $A_{1g}$ & TSE & $[z,z]$ & 0.904 & 0.751 & 0.314 \\
            $A_{2g}$ & TSE & $[y,x]-[x,y]$ & $8.65\times 10^{-3}$ & $4.69\times 10^{-2}$ & $4.11\times 10^{-2}$ \\
            $B_{1g}$ & TSE & $[y,y]-[x,x]$ & 0.227 & 1.00 & 0.745 \\
            $B_{2g}$ & TSE & $[y,x]+[x,y]$ & 0.221 & 0.329 & $7.28\times 10^{-2}$ \\
            $E_{g}$ & TSE & $\{[z,x],[z,y]\}$ & 0.404 & 0.457 & $4.00\times 10^{-2}$ \\ 
            $E_{g}$ & TSE & $\{[x,z],[y,z]\}$ & 0.163 & 0.314 & 0.235 \\ \hline
            $A_{1u}$ & SSO & $[y,y]+[x,x]$ & 0.213 & 0.672 & 0.850 \\
            $A_{1u}$ & SSO & $[z,z]$ & 0.949 & 0.795 & 0.265 \\
            $A_{2u}$ & SSO & $[y,x]-[x,y]$ & $9.36\times 10^{-2}$ & 1.00 &  1.00 \\
            $B_{1u}$ & SSO & $[y,y]-[x,x]$ & 0.172 & 0.832 & 0.912 \\
            $B_{2u}$ & SSO & $[y,x]+[x,y]$ & 0.194 & 0.967 & 0.964 \\ 
            $E_{u}$ & SSO & $\{[z,x],[z,y]\}$ & 1.13 & 1.07 & 0.428 \\
            $E_{u}$ & SSO & $\{[x,z],[y,z]\}$ & 0.144 & 0.960 & 0.748 \\ \hline
        \end{tabular}
    \end{center}
\end{table*}%
In Appendix A, we describe the three-dimensional Hamiltonian of Sr$_2$RuO$_4$ (SRO)
in the normal state in Refs.~\cite{HaverkortPRL2008,VeenstraPRL2014,RamiresPRB2019,RosisingPRR2019,SuhPRR2020,ClepkensPRR2021}.
Gell-Mann matrices $\hat{\Lambda}_{l=0\sim 8}$ are defined by
\begin{align*}
    \hat{\Lambda}_{0}&=
    \begin{pmatrix}
        1 & 0 & 0 \\
        0 & 1 & 0 \\
        0 & 0& 1
    \end{pmatrix},\hspace{1mm}
    \hat{\Lambda}_{1}=
    \begin{pmatrix}
        0 & 1 & 0 \\
        1 & 0 & 0 \\
        0 & 0& 0
    \end{pmatrix}, \\
    \hat{\Lambda}_{2}&=
    \begin{pmatrix}
        0 & 0 & 1 \\
        0 & 0 & 0 \\
        1 & 0& 0
    \end{pmatrix},\hspace{1mm}
    \hat{\Lambda}_{3}=
    \begin{pmatrix}
        0 & 0 & 0 \\
        0 & 0 & 1 \\
        0 & 1 & 0
    \end{pmatrix},\\
    \hat{\Lambda}_{4}&=
    \begin{pmatrix}
        0 & -i & 0 \\
        i & 0 & 0 \\
        0 & 0 & 0
    \end{pmatrix},\hspace{1mm}
    \hat{\Lambda}_{5}=
    \begin{pmatrix}
        0 & 0 & -i \\
        0 & 0 & 0 \\
        i & 0 & 0
    \end{pmatrix},\\
    \hat{\Lambda}_{6}&=
    \begin{pmatrix}
        0 & 0 & 0 \\
        0 & 0 & -i \\
        0 & i & 0
    \end{pmatrix},\hspace{1mm}
    \hat{\Lambda}_{7}=
    \begin{pmatrix}
        1 & 0 & 0 \\
        0 & -1 & 0 \\
        0 & 0 & 0
    \end{pmatrix},\\
    \hat{\Lambda}_{8}&=\frac{1}{\sqrt{3}}
    \begin{pmatrix}
        1 & 0 & 0 \\
        0 & 1 & 0 \\
        0 & 0 & -2
    \end{pmatrix},
\end{align*}%
in the $[d_{yz},d_{zx},d_{xy}]$ basis.
We note that the Gell-Mann matrices $\hat{\Lambda}_{l=4,5,6}$ correspond to
the $t_{2g}$-orbital angular momentum operators, 
\begin{align*}
    \hat{L}_{x}&=-\hat{\Lambda}_\mathrm{6},\hspace{1mm}
    \hat{L}_{y}=\hat{\Lambda}_\mathrm{5},\hspace{1mm}
    \hat{L}_{z}=-\hat{\Lambda}_\mathrm{4},
\end{align*}%
respectively.
The matrix elements $h_{lj}(\bm{k})$ are given by
\begin{align}
    h_{00}(\bm{k})&=\frac{1}{3}[\xi_{yz}(\bm{k})+\xi_{zx}(\bm{k})+\xi_{xy}(\bm{k})],\\
    h_{70}(\bm{k})&=\frac{1}{2}[\xi_{yz}(\bm{k})-\xi_{zx}(\bm{k})],\\
    h_{80}(\bm{k})&=\frac{1}{2\sqrt{3}}[\xi_{yz}(\bm{k})+\xi_{zx}(\bm{k})-2\xi_{xy}(\bm{k})],
\end{align}%
with intraorbital hopping terms,
\begin{align}
    h_{10}(\bm{k})&=g(\bm{k}),\\
    h_{20}(\bm{k})&=8t^{(zx,xy)}_{z}\sin\frac{k_{z}c}{2}\sin\frac{k_{x}a}{2}\cos\frac{k_{y}a}{2},\\
    h_{30}(\bm{k})&=8t^{(zx,xy)}_{z}\sin\frac{k_{z}c}{2}\cos\frac{k_{x}a}{2}\sin\frac{k_{y}a}{2},
\end{align}%
with interorbital hopping,
\begin{align}
    h_{43}(\bm{k})&=-\lambda_{z},\\
    h_{52}(\bm{k})&=-h_{61}(\bm{k})=\lambda_{xy}, 
\end{align}%
with isotropic atomic spin-orbit coupling $\lambda_{z}=\lambda_{xy}=\lambda_\mathrm{SO}$, and
\begin{align}
    h_{52}(\bm{k})&=h_{61}(\bm{k})=2\lambda^\mathrm{SOC}_{5261}[\cos{k_xa}-\cos{k_ya}],\\
    h_{51}(\bm{k})&=-h_{62}(\bm{k})=4\lambda^\mathrm{SOC}_{5162}\sin{k_xa}\sin{k_ya},\\
    h_{41}(\bm{k})&=8\lambda^\mathrm{SOC}_{12z}\sin\frac{k_{z}c}{2}\sin\frac{k_{x}a}{2}\cos\frac{k_{y}a}{2},\\
    h_{42}(\bm{k})&=8\lambda^\mathrm{SOC}_{12z}\sin\frac{k_{z}c}{2}\cos\frac{k_{x}a}{2}\sin\frac{k_{y}a}{2},\\
    h_{63}(\bm{k})&=-8\lambda^\mathrm{SOC}_{56z}\sin\frac{k_{z}c}{2}\sin\frac{k_{x}a}{2}\cos\frac{k_{y}a}{2},\\
    h_{53}(\bm{k})&=8\lambda^\mathrm{SOC}_{56z}\sin\frac{k_{z}c}{2}\cos\frac{k_{x}a}{2}\sin\frac{k_{y}a}{2},
\end{align}%
with $\bm{k}$-dependent spin-orbit coupling, respectively.
Here, $\xi_{yz,zx,xy}(\bm{k})$ and $g(\bm{k})$ are described by
\begin{align}
    \xi_{yz}(\bm{k})&=-\mu_{z}+2t^{(z,z)}_{y}\cos{k_xa}+2t^{(z,z)}_{x}\cos{k_ya}\notag\\
    &+8t^{(z,z)}_{z}\cos\frac{k_{x}a}{2}\cos\frac{k_{y}a}{2}\cos\frac{k_{z}c}{2} \notag \\
    &+4t^{(z,z)}_{xy}\cos{k_xa}\cos{k_ya}\\
    &+2t^{(z,z)}_{yy}\cos{2k_xa}+2t^{(z,z)}_{xx}\cos{2k_y}\notag \\
    &+4t^{(z,z)}_{xyy}\cos{2k_xa}\cos{k_ya}
    +4t^{(z,z)}_{xxy}\cos{2k_ya}\cos{k_xa} \notag \\
    &+2t^{(z,z)}_{zz}(\cos{k_za}-1), 
\end{align}%
\begin{align}
    \xi_{zx}(\bm{k})&=-\mu_{z}+2t^{(z,z)}_{x}\cos{k_xa}+2t^{(z,z)}_{y}\cos{k_ya}\notag\\
    &+8t^{(z,z)}_{z}\cos\frac{k_{x}a}{2}\cos\frac{k_{y}a}{2}\cos\frac{k_{z}c}{2}\notag \\
    &+4t^{(z,z)}_{xy}\cos{k_xa}\cos{k_ya}\notag\\
    &+2t^{(z,z)}_{xx}\cos{2k_xa}+2t^{(z,z)}_{yy}\cos{2k_ya}\notag \\
    &+4t^{(z,z)}_{xxy}\cos{2k_xa}\cos{k_ya}
    +4t^{(z,z)}_{xyy}\cos{2k_ya}\cos{k_xa} \notag \\
    &+2t^{(z,z)}_{zz}(\cos{k_zc}-1), 
\end{align}%
\begin{align}
    \xi_{xy}(\bm{k})&=-\mu_{xy}+2t^{(xy,xy)}_{x}(\cos{k_xa}+\cos{k_ya})\notag\\
    &+8t^{(xy,xy)}_{z}\cos\frac{k_{x}a}{2}\cos\frac{k_{y}a}{2}\cos\frac{k_{z}c}{2} \notag \\
    &+4t^{(xy,xy)}_{xy}\cos{k_xa}\cos{k_ya}\notag\\
    &+2t^{(xy,xy)}_{xx}(\cos{2k_xa}+\cos{2k_ya})\notag\\
    &+4t^{(xy,xy)}_{xxy}(\cos{2k_xa}\cos{k_ya}+\cos{2k_ya}\cos{k_xa}) \notag \\
    &+2t^{(xy,xy)}_{zz}(\cos{k_zc}-1), 
\end{align}%
\begin{align}
    g(\bm{k})&=8t^{z}_{z}\sin\frac{k_xa}{2}\sin\frac{k_ya}{2}\cos\frac{k_zc}{2}\notag\\
    &-4t^{z}_{xy}\sin{k_xa}\sin{k_ya}\notag\\
    &-4t^{z}_{xxy}(\sin{2k_xa}\sin{k_ya}+\sin{2k_ya}\sin{k_xa}).
\end{align}%
We set the parameters as shown in Table~\ref{table_A1}~\cite{SuhPRR2020}
and fix $T_\mathrm{c}=1.0\times 10^{-4}t$ with $|t^{(xy,xy)}_{x}|=t$.

\section{Orbital characters at the Fermi level in three-dimensional Sr$_2$RuO$_4$ model}

\begin{figure}[htbp]
    \centering
    \includegraphics[width=8.5cm]{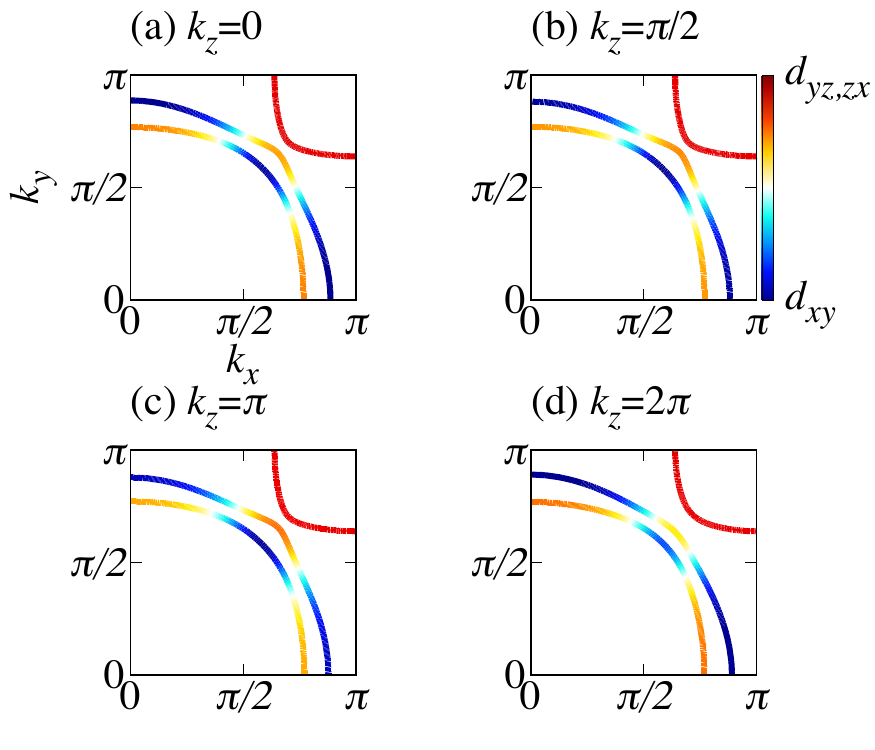}
    \caption{$t_{2g}$-orbital characters in the normal state at the Fermi level
    in (a) $k_z=0$, (b) $k_z=\pi/2$, (c) $k_z=\pi$, and (d) $k_z=2\pi$ planes. }
    \label{AFig1}
\end{figure}

Next, we confirm the orbital characters in the normal state at the Fermi level
in the three-orbital SRO model in Refs.~\cite{VeenstraPRL2014,SuhPRR2020}.
Here, we consider the density of states for each $t_{2g}$-orbital on the Ferimi surface,
\begin{align}
    &N_{\alpha}(\bm{k},E_\mathrm{F}) \notag\\
    &=-\frac{1}{\pi}\mathrm{Im}\left[G_{\alpha\uparrow,\alpha\uparrow}(\bm{k},E_\mathrm{F}+i\delta)
    +G_{\alpha\downarrow,\alpha\downarrow}(\bm{k},E_\mathrm{F}+i\delta)\right],
\end{align}%
\begin{align}
    \hat{G}(\bm{k},E_\mathrm{F})&=\frac{1}{E_\mathrm{F}+i\delta-\hat{H}(\bm{k})},
\end{align}%
with the diagonal elements of the retarded Green's function in the normal state
$G_{\alpha\uparrow,\alpha\uparrow}(\bm{k},E_\mathrm{F}+i\delta)$ and
$G_{\alpha\downarrow,\alpha\downarrow}(\bm{k},E_\mathrm{F}+i\delta)$,
$t_{2g}$-orbital indices $\alpha=yz,zx,xy$,
the Fermi energy $E_\mathrm{F}$, and the infinitesimal value $\delta$.
In Figs.~\ref{AFig1}(a), \ref{AFig1}(b), and \ref{AFig1}(c), we plot the orbital characters at the Fermi level
in Fig.~\ref{AFig1}(a) $k_z=0$, Fig.~\ref{AFig1}(b) $k_z=\pi/2$, Fig.~\ref{AFig1}(c) $k_z=\pi$, and Fig.~\ref{AFig1}(d) $k_z=2\pi$ planes
by calculating the density of states for each $t_{2g}$-orbital in the normal state.
Since the Fermi surface is cylindrical along the $k_z$-direction,
$t_{2g}$-orbital characters are almost independent of $k_z$.

\begin{figure}[htbp]
    \centering
    \includegraphics[width=8.5cm]{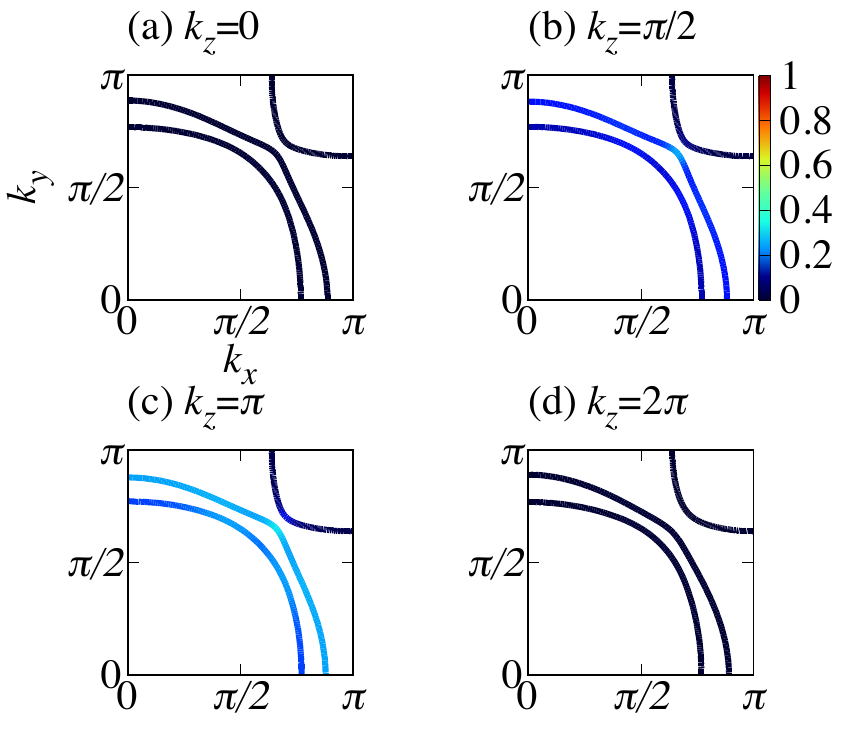}
    \caption{Gap structure of the orbital-singlet
    $E_g$ $\{[x,z],[y,z]\}$ pairing in Eq.\ (\ref{gap_Eg}) at the Fermi level 
    at (a) $k_z=0$, (b) $k_z=\pi/2$, (c) $k_z=\pi$, and (d) $k_z=2\pi$. 
    We set the temperature at $T=0$.
    Color bar indicates the gap amplitude normalized by $\Delta_{0}$. }
    \label{AFig2}
\end{figure}%
\begin{figure*}[htbp]
    \centering
    \includegraphics[width=17.5cm]{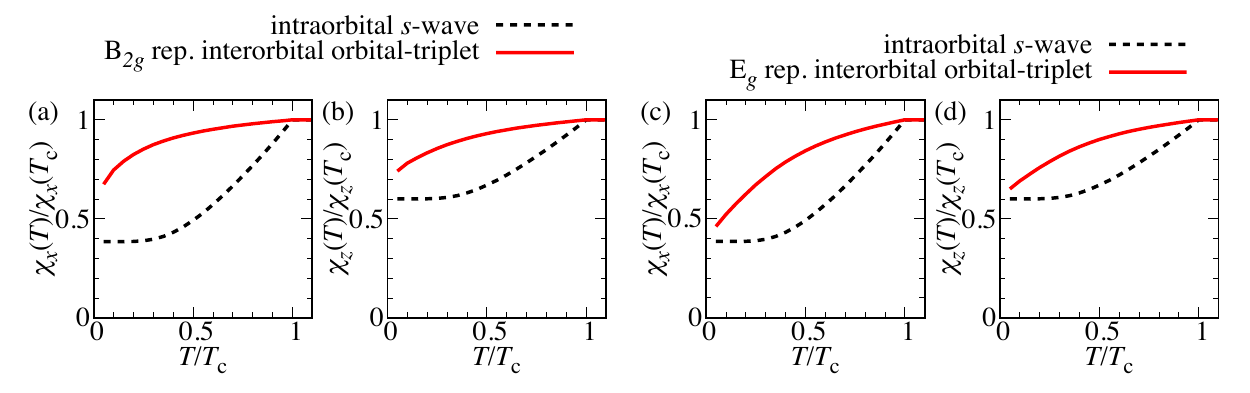}
    \caption{Spin susceptibility $\chi_{i=x,z}(T)$ for 
    interorbital STE (a)(b) $B_{2g}$ and (c)(d) $E_g$ pairings normalized by $\chi_{i}(T_\mathrm{c})$
    along the (a)(c) $x$ and (b)(d) $z$-directions
    as a function of the temperature. }
    \label{AFig3}
\end{figure*}%
\begin{figure*}[htbp]
    \centering
    \includegraphics[width=13cm]{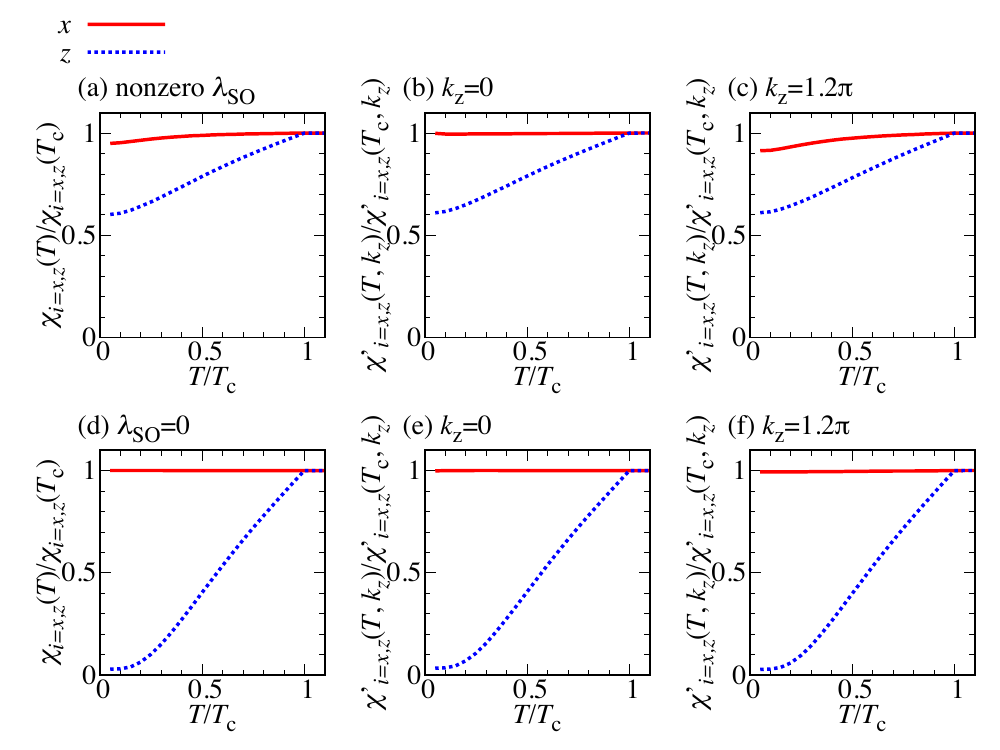}
    \caption{Spin susceptibility for the intraorbital chiral $p$-wave pairing at (a) nonzero $\lambda_\mathrm{SO}$ and (d) $\lambda_\mathrm{SO}/t=0$, normalized by $\chi_{i}(T_\mathrm{c})$
    along the $x$ (red solid line) and $z$-directions (blue dotted line)
    as a function of the temperature.
    $k_z$-resolved spin susceptibility $\chi_{i=x,z}(T)$ for 
    the intraorbital chiral $p$-wave pairing at (b)(c) nonzero $\lambda_\mathrm{SO}$ and (e)(f) $\lambda_\mathrm{SO}/t=0$ .
    We fix $k_z$ as (b)(e) $k_z=0$ and (c)(f) $k_z=1.2\pi$.
    We set the value of nonzero $\lambda_\mathrm{SO}$ as shown in Table~\ref{table_A1}.}
    \label{AFig4}
\end{figure*}%

\section{Setting of value $\alpha_c$}

Third, we summarize the constant value $\alpha_c$ for each energy band.
In SRO, the energy dispersion is described by the lowest energy band $\alpha$, $\gamma$, 
and the highest one $\beta$.
We set the constant value $\alpha_c$ as
\begin{align}
    \alpha_c&=\frac{\Delta_{0}}{\Delta_\mathrm{eff}},
\end{align}%
where $\Delta_\mathrm{eff}$ is the magnitude of the actual maximum gap amplitude 
when we set $\Delta_{0}$ in the gap function. 
Table~\ref{table_A2} shows the value of $\Delta_\mathrm{eff}/\Delta_{0}$
for each energy band, $\alpha$, $\gamma$, and $\beta$.
Since we obtain the maximum gap amplitude  
by the magnitude of the actual opening energy gap in experiments,
we modify the gap amplitude $\Delta(T)$ by using $\alpha_c$.
We choose the maximum value of $\Delta_\mathrm{eff}/\Delta_{0}$
as $\alpha_c=\Delta_0/\Delta_\mathrm{eff}$
for each irreducible representation.

\section{Gap structure of orbital-singlet $E_g$ pairing $\{[x,z],[y,z]\}$
on the Fermi surface}

In Appendix D, we confirm the gap structure in interorbital $E_g$ $\{[x,z],[y,z]\}$ pairing
on the Fermi surface.
Here, we choose the pair potential as Eq.\ (\ref{gap_Eg}).
Fig.~\ref{AFig2} shows the eigenvalues of the BdG Hamiltonian at the Fermi level
at $k_z=0$ [Fig.~\ref{AFig2}(a)], $k_z=\pi/2$ [Fig.~\ref{AFig2}(b)], $k_z=\pi$ [Fig.~\ref{AFig2}(c)], and $k_z=2\pi$ [Fig.~\ref{AFig2}(d)].
In our calculation, since we select the lower critical temperature $T_\mathrm{c}/t=1.0\times 10^{-4}$, we do not obtain the same gap structure in Ref.~\cite{SuhPRR2020}.

\section{Spin susceptibility for interorbital spin-singlet/orbital-triplet pairings}

In Appendix E, we investigate the spin susceptibility for interorbital spin-singlet/orbital-triplet/$s$-wave (STE) pairings.
The interorbital STE state appears for $B_{2g}$ and $E_{g}$ representations in Ref.~\cite{SuhPRR2020}.
Figure~\ref{AFig3} plots the spin susceptibility $\chi_{i=x,z}(T)$
as a function of the temperature along the $x$ [Figs.~\ref{AFig3}(a) and \ref{AFig3}(c)] 
and $z$-directions [Figs.~\ref{AFig3}(b) and \ref{AFig3}(d)]
for STE $B_{2g}$ [Figs.~\ref{AFig3}(a) and \ref{AFig3}(c)] 
and $E_{g}$ representations [Figs.~\ref{AFig3}(b) and \ref{AFig3}(d)].
As well as spin-triplet/orbital-singlet/$s$-wave pairings, the spin susceptibility for interorbital spin-singlet/orbital-triplet/$s$-wave pairings decreases with the temperature, independently of the axis of the applied magnetic field,
owing to the pseudospin-singlet state in the band basis. 

\section{Spin susceptibility for intraorbital chiral $p$-wave pairing}

Finally, we calculate the spin susceptibility for the intrarorbital chiral $p$-wave pairing, on behalf of all spin-triplet/odd-parity states.
The chiral $p$-wave state in the present study is given by
\begin{align}
    \hat{\Delta}(\bm{k})&=\Delta(T)\hat{L}_{0}\otimes\hat{\sigma}_{z}[\sin{k_x}+i\sin{k_y}]i\hat{\sigma}_{y},
\end{align}%
with the unit matrix in $t_{2g}$-orbital space $\hat{L}_{0}$. 

Figure~\ref{AFig4} plots the spin susceptibility $\chi_{i=x,z}(T)$ for the intraorbital chiral $p$-wave state as a function of the temperature at nonzero $\lambda_\mathrm{SO}$ in Table~\ref{table_A1} [Figs.~\ref{AFig4}(a) to \ref{AFig4}(c)] and $\lambda_\mathrm{SO}/t=0$
[Figs.~\ref{AFig4}(d) to \ref{AFig4}(f)].
It includes the $k_z$-resolved spin susceptibility $\chi{'}_{i=x,z}(T,k_z)$ at (b)(e) $k_z=0$
[Figs.~\ref{AFig4}(b) and \ref{AFig4}(e)] and $k_z=1.2\pi$ [Figs.~\ref{AFig4}(c) and \ref{AFig4}(f)].
Here, the spin susceptibility $\chi_{i=x,z}(T)$ is described by
\begin{align}
    \chi_{i}(T)\sim\int^{2\pi}_{-2\pi}\chi{'}_{i}(T,k_z)dk_{z}.
\end{align}%
The spin susceptibility at nonzero atomic spin-orbit coupling $\lambda_\mathrm{SO}$
decreases around $5\%$ along the $x$-direction at low temperature as shown in Fig.~\ref{AFig4}(a).
To analyze this behavior, we resolve the spin susceptibility for $k_z$. 
Then we select $k_z=0$ and $k_z=1.2\pi$.
Because the $k_z$-resolved spin susceptibility along the $x$-axis changes remarkably when $k_z$ is larger than $\pi$, we choose $k_z=1.2\pi$ in Fig.~\ref{AFig4}.
At $k_z=0$, the $k_z$-resolved spin susceptibility does not change along the $x$-direction 
as shown in Fig.~\ref{AFig4}(b). 
As $k_z=0$ ($k_z=2\pi$) is on the symmetric line (the edge of the Brillouin zone), the pseudospin \textbf{d}-vector should be aligned along the $x$-axis in the intraorbital chiral $p$-wave pairing.
However, the $k_z$-resolved spin susceptibility at $k_z=1.2\pi$ decreases around $10\%$ along the $x$-axis as shown in Fig.~\ref{AFig4}(c).
Thus, the spin susceptibility is reduced along the $x$-direction by the components away from the $xy$-symmetric plane.

To unveil the role of the atomic spin-orbit coupling $\lambda_\mathrm{SO}$, we also study the spin susceptibility at $\lambda_\mathrm{SO}/t=0$.
At $\lambda_\mathrm{SO}/t=0$, the spin susceptibility does not decrease along the $x$-direction in Fig.~\ref{AFig4}(d). 
Since the pseudospin \textbf{d}-vector for the chiral $p$-wave pairing is completely aligned along the $z$-axis for all $k_z$, the $k_z$-resolved spin susceptibility at both $k_z=0$ and $1.2\pi$ does not change along the $x$-axis as shown in Figs.~\ref{AFig4}(e) and \ref{AFig4}(f). 

In conclusion, at the nonzero atomic spin-orbit coupling $\lambda_\mathrm{SO}$, when we can define the pseudospin \textbf{d}-vector, the spin susceptibility is reduced around $5$-$10\%$ along the axis where there is no reduction in the single-orbital model.
It occurs by the pseudospin \textbf{d}-vector that is not completely aligned in the $xy$-plane or $z$-direction away from the $xy$-symmetric plane in the presence of the strong atomic spin-orbit coupling $\lambda_\mathrm{SO}$.

\bibliography{main}

\begin{thebibliography}{70}%
\makeatletter
\providecommand \@ifxundefined [1]{%
 \@ifx{#1\undefined}
}%
\providecommand \@ifnum [1]{%
 \ifnum #1\expandafter \@firstoftwo
 \else \expandafter \@secondoftwo
 \fi
}%
\providecommand \@ifx [1]{%
 \ifx #1\expandafter \@firstoftwo
 \else \expandafter \@secondoftwo
 \fi
}%
\providecommand \natexlab [1]{#1}%
\providecommand \enquote  [1]{``#1''}%
\providecommand \bibnamefont  [1]{#1}%
\providecommand \bibfnamefont [1]{#1}%
\providecommand \citenamefont [1]{#1}%
\providecommand \href@noop [0]{\@secondoftwo}%
\providecommand \href [0]{\begingroup \@sanitize@url \@href}%
\providecommand \@href[1]{\@@startlink{#1}\@@href}%
\providecommand \@@href[1]{\endgroup#1\@@endlink}%
\providecommand \@sanitize@url [0]{\catcode `\\12\catcode `\$12\catcode
  `\&12\catcode `\#12\catcode `\^12\catcode `\_12\catcode `\%12\relax}%
\providecommand \@@startlink[1]{}%
\providecommand \@@endlink[0]{}%
\providecommand \url  [0]{\begingroup\@sanitize@url \@url }%
\providecommand \@url [1]{\endgroup\@href {#1}{\urlprefix }}%
\providecommand \urlprefix  [0]{URL }%
\providecommand \Eprint [0]{\href }%
\providecommand \doibase [0]{https://doi.org/}%
\providecommand \selectlanguage [0]{\@gobble}%
\providecommand \bibinfo  [0]{\@secondoftwo}%
\providecommand \bibfield  [0]{\@secondoftwo}%
\providecommand \translation [1]{[#1]}%
\providecommand \BibitemOpen [0]{}%
\providecommand \bibitemStop [0]{}%
\providecommand \bibitemNoStop [0]{.\EOS\space}%
\providecommand \EOS [0]{\spacefactor3000\relax}%
\providecommand \BibitemShut  [1]{\csname bibitem#1\endcsname}%
\let\auto@bib@innerbib\@empty
\bibitem [{\citenamefont {Maeno}\ \emph {et~al.}(1994)\citenamefont {Maeno},
  \citenamefont {Hashimoto}, \citenamefont {Yoshida}, \citenamefont
  {Nishizaki}, \citenamefont {Fujita}, \citenamefont {Bednorz},\ and\
  \citenamefont {Lichtenberg}}]{Maeno}%
  \BibitemOpen
  \bibfield  {author} {\bibinfo {author} {\bibfnamefont {Y.}~\bibnamefont
  {Maeno}}, \bibinfo {author} {\bibfnamefont {H.}~\bibnamefont {Hashimoto}},
  \bibinfo {author} {\bibfnamefont {K.}~\bibnamefont {Yoshida}}, \bibinfo
  {author} {\bibfnamefont {S.}~\bibnamefont {Nishizaki}}, \bibinfo {author}
  {\bibfnamefont {T.}~\bibnamefont {Fujita}}, \bibinfo {author} {\bibfnamefont
  {J.~G.}\ \bibnamefont {Bednorz}},\ and\ \bibinfo {author} {\bibfnamefont
  {F.}~\bibnamefont {Lichtenberg}},\ }\bibfield  {title} {\bibinfo {title}
  {Superconductivity in a layered perovskite without copper},\ }\href@noop {}
  {\bibfield  {journal} {\bibinfo  {journal} {Nature}\ }\textbf {\bibinfo
  {volume} {372}},\ \bibinfo {pages} {532} (\bibinfo {year}
  {1994})}\BibitemShut {NoStop}%
\bibitem [{\citenamefont {Mackenzie}\ and\ \citenamefont
  {Maeno}(2003)}]{MackenzieRMP2003}%
  \BibitemOpen
  \bibfield  {author} {\bibinfo {author} {\bibfnamefont {A.~P.}\ \bibnamefont
  {Mackenzie}}\ and\ \bibinfo {author} {\bibfnamefont {Y.}~\bibnamefont
  {Maeno}},\ }\bibfield  {title} {\bibinfo {title} {The superconductivity of
  ${\mathrm{sr}}_{2}{\mathrm{ruo}}_{4}$ and the physics of spin-triplet
  pairing},\ }\href {https://doi.org/10.1103/RevModPhys.75.657} {\bibfield
  {journal} {\bibinfo  {journal} {Rev. Mod. Phys.}\ }\textbf {\bibinfo {volume}
  {75}},\ \bibinfo {pages} {657} (\bibinfo {year} {2003})}\BibitemShut
  {NoStop}%
\bibitem [{\citenamefont {Maeno}\ \emph {et~al.}(2012)\citenamefont {Maeno},
  \citenamefont {Kittaka}, \citenamefont {Nomura}, \citenamefont {Yonezawa},\
  and\ \citenamefont {Ishida}}]{MaenoJPSJ2012}%
  \BibitemOpen
  \bibfield  {author} {\bibinfo {author} {\bibfnamefont {Y.}~\bibnamefont
  {Maeno}}, \bibinfo {author} {\bibfnamefont {S.}~\bibnamefont {Kittaka}},
  \bibinfo {author} {\bibfnamefont {T.}~\bibnamefont {Nomura}}, \bibinfo
  {author} {\bibfnamefont {S.}~\bibnamefont {Yonezawa}},\ and\ \bibinfo
  {author} {\bibfnamefont {K.}~\bibnamefont {Ishida}},\ }\bibfield  {title}
  {\bibinfo {title} {Evaluation of spin-triplet superconductivity in
  ${\mathrm{sr}}_{2}{\mathrm{ruo}}_{4}$},\ }\href
  {https://doi.org/10.1143/JPSJ.81.011009} {\bibfield  {journal} {\bibinfo
  {journal} {J. Phys. Soc. Jpn.}\ }\textbf {\bibinfo {volume} {81}},\ \bibinfo
  {pages} {011009} (\bibinfo {year} {2012})}\BibitemShut {NoStop}%
\bibitem [{\citenamefont {Duffy}\ \emph {et~al.}(2000)\citenamefont {Duffy},
  \citenamefont {Hayden}, \citenamefont {Maeno}, \citenamefont {Mao},
  \citenamefont {Kulda},\ and\ \citenamefont {McIntyre}}]{DuffyPRL2000}%
  \BibitemOpen
  \bibfield  {author} {\bibinfo {author} {\bibfnamefont {J.~A.}\ \bibnamefont
  {Duffy}}, \bibinfo {author} {\bibfnamefont {S.~M.}\ \bibnamefont {Hayden}},
  \bibinfo {author} {\bibfnamefont {Y.}~\bibnamefont {Maeno}}, \bibinfo
  {author} {\bibfnamefont {Z.}~\bibnamefont {Mao}}, \bibinfo {author}
  {\bibfnamefont {J.}~\bibnamefont {Kulda}},\ and\ \bibinfo {author}
  {\bibfnamefont {G.~J.}\ \bibnamefont {McIntyre}},\ }\bibfield  {title}
  {\bibinfo {title} {Polarized-neutron scattering study of the cooper-pair
  moment in ${\mathrm{sr}}_{2}{\mathrm{ruo}}_{4}$},\ }\href
  {https://doi.org/10.1103/PhysRevLett.85.5412} {\bibfield  {journal} {\bibinfo
   {journal} {Phys. Rev. Lett.}\ }\textbf {\bibinfo {volume} {85}},\ \bibinfo
  {pages} {5412} (\bibinfo {year} {2000})}\BibitemShut {NoStop}%
\bibitem [{\citenamefont {Jang}\ \emph {et~al.}(2011)\citenamefont {Jang},
  \citenamefont {Ferguson}, \citenamefont {Vakaryuk}, \citenamefont {Budakian},
  \citenamefont {Chung}, \citenamefont {Goldbart},\ and\ \citenamefont
  {Maeno}}]{JangScience2011}%
  \BibitemOpen
  \bibfield  {author} {\bibinfo {author} {\bibfnamefont {J.}~\bibnamefont
  {Jang}}, \bibinfo {author} {\bibfnamefont {D.}~\bibnamefont {Ferguson}},
  \bibinfo {author} {\bibfnamefont {V.}~\bibnamefont {Vakaryuk}}, \bibinfo
  {author} {\bibfnamefont {R.}~\bibnamefont {Budakian}}, \bibinfo {author}
  {\bibfnamefont {S.}~\bibnamefont {Chung}}, \bibinfo {author} {\bibfnamefont
  {P.}~\bibnamefont {Goldbart}},\ and\ \bibinfo {author} {\bibfnamefont
  {Y.}~\bibnamefont {Maeno}},\ }\bibfield  {title} {\bibinfo {title}
  {Observation of half-height magnetization steps in
  ${\mathrm{sr}}_{2}{\mathrm{ruo}}_{4}$},\ }\href@noop {} {\bibfield  {journal}
  {\bibinfo  {journal} {Science}\ }\textbf {\bibinfo {volume} {331}},\ \bibinfo
  {pages} {186} (\bibinfo {year} {2011})}\BibitemShut {NoStop}%
\bibitem [{\citenamefont {Yasui}\ \emph {et~al.}(2017)\citenamefont {Yasui},
  \citenamefont {Lahabi}, \citenamefont {Anwar}, \citenamefont {Nakamura},
  \citenamefont {Yonezawa}, \citenamefont {Terashima}, \citenamefont {Aarts},\
  and\ \citenamefont {Maeno}}]{YasuiPRB2017}%
  \BibitemOpen
  \bibfield  {author} {\bibinfo {author} {\bibfnamefont {Y.}~\bibnamefont
  {Yasui}}, \bibinfo {author} {\bibfnamefont {K.}~\bibnamefont {Lahabi}},
  \bibinfo {author} {\bibfnamefont {M.~S.}\ \bibnamefont {Anwar}}, \bibinfo
  {author} {\bibfnamefont {Y.}~\bibnamefont {Nakamura}}, \bibinfo {author}
  {\bibfnamefont {S.}~\bibnamefont {Yonezawa}}, \bibinfo {author}
  {\bibfnamefont {T.}~\bibnamefont {Terashima}}, \bibinfo {author}
  {\bibfnamefont {J.}~\bibnamefont {Aarts}},\ and\ \bibinfo {author}
  {\bibfnamefont {Y.}~\bibnamefont {Maeno}},\ }\bibfield  {title} {\bibinfo
  {title} {Little-parks oscillations with half-quantum fluxoid features in
  ${\mathrm{sr}}_{2}{\mathrm{ruo}}_{4}$ microrings},\ }\href@noop {} {\bibfield
   {journal} {\bibinfo  {journal} {Phys. Rev. B}\ }\textbf {\bibinfo {volume}
  {96}},\ \bibinfo {pages} {180507(R)} (\bibinfo {year} {2017})}\BibitemShut
  {NoStop}%
\bibitem [{\citenamefont {Yamashiro}\ \emph {et~al.}(1998)\citenamefont
  {Yamashiro}, \citenamefont {Tanaka},\ and\ \citenamefont
  {Kashiwaya}}]{YamashiroJPSJ}%
  \BibitemOpen
  \bibfield  {author} {\bibinfo {author} {\bibfnamefont {M.}~\bibnamefont
  {Yamashiro}}, \bibinfo {author} {\bibfnamefont {Y.}~\bibnamefont {Tanaka}},\
  and\ \bibinfo {author} {\bibfnamefont {S.}~\bibnamefont {Kashiwaya}},\
  }\bibfield  {title} {\bibinfo {title} {Theory of the d.c. josephson effect in
  $s$-wave/$p$-wave/$s$-wave superconductor junction},\ }\href@noop {}
  {\bibfield  {journal} {\bibinfo  {journal} {J. Phys. Soc. Jpn.}\ }\textbf
  {\bibinfo {volume} {67}},\ \bibinfo {pages} {3364} (\bibinfo {year}
  {1998})}\BibitemShut {NoStop}%
\bibitem [{\citenamefont {Jin}\ \emph {et~al.}(1999)\citenamefont {Jin},
  \citenamefont {Zadorozhny}, \citenamefont {Liu}, \citenamefont {Schlom},
  \citenamefont {Mori},\ and\ \citenamefont {Maeno}}]{JinPRB1999}%
  \BibitemOpen
  \bibfield  {author} {\bibinfo {author} {\bibfnamefont {R.}~\bibnamefont
  {Jin}}, \bibinfo {author} {\bibfnamefont {Y.}~\bibnamefont {Zadorozhny}},
  \bibinfo {author} {\bibfnamefont {Y.}~\bibnamefont {Liu}}, \bibinfo {author}
  {\bibfnamefont {D.~G.}\ \bibnamefont {Schlom}}, \bibinfo {author}
  {\bibfnamefont {Y.}~\bibnamefont {Mori}},\ and\ \bibinfo {author}
  {\bibfnamefont {Y.}~\bibnamefont {Maeno}},\ }\bibfield  {title} {\bibinfo
  {title} {Observation of anomalous temperature dependence of the critical
  current in
  ${\mathrm{p}\mathrm{b}/\mathrm{s}\mathrm{r}}_{2}{\mathrm{ruo}}_{4}/\mathrm{Pb}$
  junctions},\ }\href {https://doi.org/10.1103/PhysRevB.59.4433} {\bibfield
  {journal} {\bibinfo  {journal} {Phys. Rev. B}\ }\textbf {\bibinfo {volume}
  {59}},\ \bibinfo {pages} {4433} (\bibinfo {year} {1999})}\BibitemShut
  {NoStop}%
\bibitem [{\citenamefont {Laube}\ \emph {et~al.}(2000)\citenamefont {Laube},
  \citenamefont {Goll}, \citenamefont {L\"ohneysen}, \citenamefont
  {Fogelstr\"om},\ and\ \citenamefont {Lichtenberg}}]{LaubePRL2000}%
  \BibitemOpen
  \bibfield  {author} {\bibinfo {author} {\bibfnamefont {F.}~\bibnamefont
  {Laube}}, \bibinfo {author} {\bibfnamefont {G.}~\bibnamefont {Goll}},
  \bibinfo {author} {\bibfnamefont {H.~v.}\ \bibnamefont {L\"ohneysen}},
  \bibinfo {author} {\bibfnamefont {M.}~\bibnamefont {Fogelstr\"om}},\ and\
  \bibinfo {author} {\bibfnamefont {F.}~\bibnamefont {Lichtenberg}},\
  }\bibfield  {title} {\bibinfo {title} {Spin-triplet superconductivity in
  ${\mathrm{sr}}_{2}{\mathrm{ruo}}_{4}$ probed by andreev reflection},\ }\href
  {https://doi.org/10.1103/PhysRevLett.84.1595} {\bibfield  {journal} {\bibinfo
   {journal} {Phys. Rev. Lett.}\ }\textbf {\bibinfo {volume} {84}},\ \bibinfo
  {pages} {1595} (\bibinfo {year} {2000})}\BibitemShut {NoStop}%
\bibitem [{\citenamefont {Tanaka}\ \emph {et~al.}(2009)\citenamefont {Tanaka},
  \citenamefont {Yokoyama}, \citenamefont {Balatsky},\ and\ \citenamefont
  {Nagaosa}}]{Tanaka2009}%
  \BibitemOpen
  \bibfield  {author} {\bibinfo {author} {\bibfnamefont {Y.}~\bibnamefont
  {Tanaka}}, \bibinfo {author} {\bibfnamefont {T.}~\bibnamefont {Yokoyama}},
  \bibinfo {author} {\bibfnamefont {A.~V.}\ \bibnamefont {Balatsky}},\ and\
  \bibinfo {author} {\bibfnamefont {N.}~\bibnamefont {Nagaosa}},\ }\bibfield
  {title} {\bibinfo {title} {Theory of topological spin current in
  noncentrosymmetric superconductors},\ }\href@noop {} {\bibfield  {journal}
  {\bibinfo  {journal} {Phys. Rev. B}\ }\textbf {\bibinfo {volume} {79}},\
  \bibinfo {pages} {060505(R)} (\bibinfo {year} {2009})}\BibitemShut {NoStop}%
\bibitem [{\citenamefont {Wu}\ and\ \citenamefont
  {Samokhin}(2010)}]{WuPRB2010}%
  \BibitemOpen
  \bibfield  {author} {\bibinfo {author} {\bibfnamefont {S.}~\bibnamefont
  {Wu}}\ and\ \bibinfo {author} {\bibfnamefont {K.~V.}\ \bibnamefont
  {Samokhin}},\ }\bibfield  {title} {\bibinfo {title} {Effects of interface
  spin-orbit coupling on tunneling between normal metal and chiral $p$-wave
  superconductor},\ }\href {https://doi.org/10.1103/PhysRevB.81.214506}
  {\bibfield  {journal} {\bibinfo  {journal} {Phys. Rev. B}\ }\textbf {\bibinfo
  {volume} {81}},\ \bibinfo {pages} {214506} (\bibinfo {year}
  {2010})}\BibitemShut {NoStop}%
\bibitem [{\citenamefont {Kashiwaya}\ \emph {et~al.}(2011)\citenamefont
  {Kashiwaya}, \citenamefont {Kashiwaya}, \citenamefont {Kambara},
  \citenamefont {Furuta}, \citenamefont {Yaguchi}, \citenamefont {Tanaka},\
  and\ \citenamefont {Maeno}}]{KashiwayaPRL2011}%
  \BibitemOpen
  \bibfield  {author} {\bibinfo {author} {\bibfnamefont {S.}~\bibnamefont
  {Kashiwaya}}, \bibinfo {author} {\bibfnamefont {H.}~\bibnamefont
  {Kashiwaya}}, \bibinfo {author} {\bibfnamefont {H.}~\bibnamefont {Kambara}},
  \bibinfo {author} {\bibfnamefont {T.}~\bibnamefont {Furuta}}, \bibinfo
  {author} {\bibfnamefont {H.}~\bibnamefont {Yaguchi}}, \bibinfo {author}
  {\bibfnamefont {Y.}~\bibnamefont {Tanaka}},\ and\ \bibinfo {author}
  {\bibfnamefont {Y.}~\bibnamefont {Maeno}},\ }\bibfield  {title} {\bibinfo
  {title} {Edge states of ${\mathrm{sr}}_{2}{\mathrm{ruo}}_{4}$ detected by
  in-plane tunneling spectroscopy},\ }\href@noop {} {\bibfield  {journal}
  {\bibinfo  {journal} {Phys. Rev. Lett.}\ }\textbf {\bibinfo {volume} {107}},\
  \bibinfo {pages} {077003} (\bibinfo {year} {2011})}\BibitemShut {NoStop}%
\bibitem [{\citenamefont {Anwar}\ \emph {et~al.}(2016)\citenamefont {Anwar},
  \citenamefont {Lee}, \citenamefont {Ishiguro}, \citenamefont {Sugimoto},
  \citenamefont {Tano}, \citenamefont {Kang}, \citenamefont {Shin},
  \citenamefont {Yonezawa}, \citenamefont {Manske}, \citenamefont {Takayanagi},
  \citenamefont {Noh},\ and\ \citenamefont {Maeno}}]{AnwarNatPhys2016}%
  \BibitemOpen
  \bibfield  {author} {\bibinfo {author} {\bibfnamefont {M.}~\bibnamefont
  {Anwar}}, \bibinfo {author} {\bibfnamefont {S.}~\bibnamefont {Lee}}, \bibinfo
  {author} {\bibfnamefont {R.}~\bibnamefont {Ishiguro}}, \bibinfo {author}
  {\bibfnamefont {Y.}~\bibnamefont {Sugimoto}}, \bibinfo {author}
  {\bibfnamefont {Y.}~\bibnamefont {Tano}}, \bibinfo {author} {\bibfnamefont
  {S.}~\bibnamefont {Kang}}, \bibinfo {author} {\bibfnamefont {Y.}~\bibnamefont
  {Shin}}, \bibinfo {author} {\bibfnamefont {S.}~\bibnamefont {Yonezawa}},
  \bibinfo {author} {\bibfnamefont {D.}~\bibnamefont {Manske}}, \bibinfo
  {author} {\bibfnamefont {H.}~\bibnamefont {Takayanagi}}, \bibinfo {author}
  {\bibfnamefont {T.~W.}\ \bibnamefont {Noh}},\ and\ \bibinfo {author}
  {\bibfnamefont {Y.}~\bibnamefont {Maeno}},\ }\bibfield  {title} {\bibinfo
  {title} {Direct penetration of spin-triplet superconductivity into a
  ferromagnet in
  au/${\mathrm{sr}}{\mathrm{ruo}}_{3}$/${\mathrm{sr}}_{2}{\mathrm{ruo}}_{4}$
  junctions},\ }\href@noop {} {\bibfield  {journal} {\bibinfo  {journal} {Nat.
  Commun.}\ }\textbf {\bibinfo {volume} {7}},\ \bibinfo {pages} {13220}
  (\bibinfo {year} {2016})}\BibitemShut {NoStop}%
\bibitem [{\citenamefont {Olde~Olthof}\ \emph {et~al.}(2018)\citenamefont
  {Olde~Olthof}, \citenamefont {Suzuki}, \citenamefont {Golubov}, \citenamefont
  {Kunieda}, \citenamefont {Yonezawa}, \citenamefont {Maeno},\ and\
  \citenamefont {Tanaka}}]{OldePRB2018}%
  \BibitemOpen
  \bibfield  {author} {\bibinfo {author} {\bibfnamefont {L.~A.~B.}\
  \bibnamefont {Olde~Olthof}}, \bibinfo {author} {\bibfnamefont {S.-I.}\
  \bibnamefont {Suzuki}}, \bibinfo {author} {\bibfnamefont {A.~A.}\
  \bibnamefont {Golubov}}, \bibinfo {author} {\bibfnamefont {M.}~\bibnamefont
  {Kunieda}}, \bibinfo {author} {\bibfnamefont {S.}~\bibnamefont {Yonezawa}},
  \bibinfo {author} {\bibfnamefont {Y.}~\bibnamefont {Maeno}},\ and\ \bibinfo
  {author} {\bibfnamefont {Y.}~\bibnamefont {Tanaka}},\ }\bibfield  {title}
  {\bibinfo {title} {Theory of tunneling spectroscopy of normal
  metal/ferromagnet/spin-triplet superconductor junctions},\ }\href
  {https://doi.org/10.1103/PhysRevB.98.014508} {\bibfield  {journal} {\bibinfo
  {journal} {Phys. Rev. B}\ }\textbf {\bibinfo {volume} {98}},\ \bibinfo
  {pages} {014508} (\bibinfo {year} {2018})}\BibitemShut {NoStop}%
\bibitem [{\citenamefont {Ishida}\ \emph {et~al.}(1998)\citenamefont {Ishida},
  \citenamefont {Mukuda}, \citenamefont {Kitaoka}, \citenamefont {Asayama},
  \citenamefont {Mao}, \citenamefont {Mori},\ and\ \citenamefont
  {Maeno}}]{IshidaNature1998}%
  \BibitemOpen
  \bibfield  {author} {\bibinfo {author} {\bibfnamefont {K.}~\bibnamefont
  {Ishida}}, \bibinfo {author} {\bibfnamefont {H.}~\bibnamefont {Mukuda}},
  \bibinfo {author} {\bibfnamefont {Y.}~\bibnamefont {Kitaoka}}, \bibinfo
  {author} {\bibfnamefont {K.}~\bibnamefont {Asayama}}, \bibinfo {author}
  {\bibfnamefont {Z.~Q.}\ \bibnamefont {Mao}}, \bibinfo {author} {\bibfnamefont
  {Y.}~\bibnamefont {Mori}},\ and\ \bibinfo {author} {\bibfnamefont
  {Y.}~\bibnamefont {Maeno}},\ }\bibfield  {title} {\bibinfo {title}
  {Spin-triplet superconductivity in ${\mathrm{sr}}_{2}{\mathrm{ruo}}_{4}$
  identified by 17o knight shift},\ }\href {https://doi.org/10.1038/25315}
  {\bibfield  {journal} {\bibinfo  {journal} {Nature}\ }\textbf {\bibinfo
  {volume} {396}},\ \bibinfo {pages} {658} (\bibinfo {year}
  {1998})}\BibitemShut {NoStop}%
\bibitem [{\citenamefont {Rice}\ and\ \citenamefont
  {Sigrist}(1995)}]{RiceJPCM1995}%
  \BibitemOpen
  \bibfield  {author} {\bibinfo {author} {\bibfnamefont {T.~M.}\ \bibnamefont
  {Rice}}\ and\ \bibinfo {author} {\bibfnamefont {M.}~\bibnamefont {Sigrist}},\
  }\bibfield  {title} {\bibinfo {title} {An electronic analogue of 3he?},\
  }\href@noop {} {\bibfield  {journal} {\bibinfo  {journal} {J. Phys.: Condens.
  Matter}\ }\textbf {\bibinfo {volume} {7}},\ \bibinfo {pages} {L643} (\bibinfo
  {year} {1995})}\BibitemShut {NoStop}%
\bibitem [{\citenamefont {Nomura}\ and\ \citenamefont
  {Yamada}(2000)}]{NomuraJPSJ2000}%
  \BibitemOpen
  \bibfield  {author} {\bibinfo {author} {\bibfnamefont {T.}~\bibnamefont
  {Nomura}}\ and\ \bibinfo {author} {\bibfnamefont {K.}~\bibnamefont
  {Yamada}},\ }\bibfield  {title} {\bibinfo {title} {Perturbation theory of
  spin-triplet superconductivity for ${\mathrm{sr}}_{2}{\mathrm{ruo}}_{4}$},\
  }\href@noop {} {\bibfield  {journal} {\bibinfo  {journal} {J. Phys. Soc.
  Jpn.}\ }\textbf {\bibinfo {volume} {69}},\ \bibinfo {pages} {3678} (\bibinfo
  {year} {2000})}\BibitemShut {NoStop}%
\bibitem [{\citenamefont {Sato}\ and\ \citenamefont
  {Kohmoto}(2000)}]{SatoJPSJ2000}%
  \BibitemOpen
  \bibfield  {author} {\bibinfo {author} {\bibfnamefont {M.}~\bibnamefont
  {Sato}}\ and\ \bibinfo {author} {\bibfnamefont {M.}~\bibnamefont {Kohmoto}},\
  }\bibfield  {title} {\bibinfo {title} {Mechanism of spin-triplet
  superconductivity in ${\mathrm{sr}}_{2}{\mathrm{ruo}}_{4}$},\ }\href@noop {}
  {\bibfield  {journal} {\bibinfo  {journal} {J. Phys. Soc. Jpn.}\ }\textbf
  {\bibinfo {volume} {69}},\ \bibinfo {pages} {3505} (\bibinfo {year}
  {2000})}\BibitemShut {NoStop}%
\bibitem [{\citenamefont {Takimoto}(2000)}]{TakimotoPRB2000}%
  \BibitemOpen
  \bibfield  {author} {\bibinfo {author} {\bibfnamefont {T.}~\bibnamefont
  {Takimoto}},\ }\bibfield  {title} {\bibinfo {title} {Orbital
  fluctuation-induced triplet superconductivity: Mechanism of superconductivity
  in ${\mathrm{sr}}_{2}{\mathrm{ruo}}_{4}$},\ }\href
  {https://doi.org/10.1103/PhysRevB.62.R14641} {\bibfield  {journal} {\bibinfo
  {journal} {Phys. Rev. B}\ }\textbf {\bibinfo {volume} {62}},\ \bibinfo
  {pages} {R14641} (\bibinfo {year} {2000})}\BibitemShut {NoStop}%
\bibitem [{\citenamefont {Kuroki}\ \emph {et~al.}(2001)\citenamefont {Kuroki},
  \citenamefont {Ogata}, \citenamefont {Arita},\ and\ \citenamefont
  {Aoki}}]{KurokiPRB2001}%
  \BibitemOpen
  \bibfield  {author} {\bibinfo {author} {\bibfnamefont {K.}~\bibnamefont
  {Kuroki}}, \bibinfo {author} {\bibfnamefont {M.}~\bibnamefont {Ogata}},
  \bibinfo {author} {\bibfnamefont {R.}~\bibnamefont {Arita}},\ and\ \bibinfo
  {author} {\bibfnamefont {H.}~\bibnamefont {Aoki}},\ }\bibfield  {title}
  {\bibinfo {title} {Crib-shaped triplet-pairing gap function for an orthogonal
  pair of quasi-one-dimensional fermi surfaces in
  ${\mathrm{sr}}_{2}{\mathrm{ruo}}_{4}$},\ }\href
  {https://doi.org/10.1103/PhysRevB.63.060506} {\bibfield  {journal} {\bibinfo
  {journal} {Phys. Rev. B}\ }\textbf {\bibinfo {volume} {63}},\ \bibinfo
  {pages} {060506} (\bibinfo {year} {2001})}\BibitemShut {NoStop}%
\bibitem [{\citenamefont {Nomura}\ and\ \citenamefont
  {Yamada}(2002{\natexlab{a}})}]{NomuraJPSJ2002}%
  \BibitemOpen
  \bibfield  {author} {\bibinfo {author} {\bibfnamefont {T.}~\bibnamefont
  {Nomura}}\ and\ \bibinfo {author} {\bibfnamefont {K.}~\bibnamefont
  {Yamada}},\ }\bibfield  {title} {\bibinfo {title} {Detailed investigation of
  gap structure and specific heat in the $p$-wave superconductor
  ${\mathrm{sr}}_{2}{\mathrm{ruo}}_{4}$},\ }\href@noop {} {\bibfield  {journal}
  {\bibinfo  {journal} {J. Phys. Soc. Jpn.}\ }\textbf {\bibinfo {volume}
  {71}},\ \bibinfo {pages} {404} (\bibinfo {year}
  {2002}{\natexlab{a}})}\BibitemShut {NoStop}%
\bibitem [{\citenamefont {Nomura}\ and\ \citenamefont
  {Yamada}(2002{\natexlab{b}})}]{NomuraJPSJ2002_2}%
  \BibitemOpen
  \bibfield  {author} {\bibinfo {author} {\bibfnamefont {T.}~\bibnamefont
  {Nomura}}\ and\ \bibinfo {author} {\bibfnamefont {K.}~\bibnamefont
  {Yamada}},\ }\bibfield  {title} {\bibinfo {title} {Roles of electron
  correlations in the spin-triplet superconductivity of
  ${\mathrm{sr}}_{2}{\mathrm{ruo}}_{4}$},\ }\href@noop {} {\bibfield  {journal}
  {\bibinfo  {journal} {J. Phys. Soc. Jpn.}\ }\textbf {\bibinfo {volume}
  {71}},\ \bibinfo {pages} {1993} (\bibinfo {year}
  {2002}{\natexlab{b}})}\BibitemShut {NoStop}%
\bibitem [{\citenamefont {Yanase}\ and\ \citenamefont
  {Ogata}(2003)}]{YanaseJPSJ2003}%
  \BibitemOpen
  \bibfield  {author} {\bibinfo {author} {\bibfnamefont {Y.}~\bibnamefont
  {Yanase}}\ and\ \bibinfo {author} {\bibfnamefont {M.}~\bibnamefont {Ogata}},\
  }\bibfield  {title} {\bibinfo {title} {Microscopic identification of the
  $d$-vector in triplet superconductor ${\mathrm{sr}}_{2}{\mathrm{ruo}}_{4}$},\
  }\href@noop {} {\bibfield  {journal} {\bibinfo  {journal} {J. Phys. Soc.
  Jpn.}\ }\textbf {\bibinfo {volume} {72}},\ \bibinfo {pages} {673} (\bibinfo
  {year} {2003})}\BibitemShut {NoStop}%
\bibitem [{\citenamefont {Nomura}\ and\ \citenamefont
  {Yamada}(2005)}]{NomuraJPSJ2005}%
  \BibitemOpen
  \bibfield  {author} {\bibinfo {author} {\bibfnamefont {T.}~\bibnamefont
  {Nomura}}\ and\ \bibinfo {author} {\bibfnamefont {K.}~\bibnamefont
  {Yamada}},\ }\bibfield  {title} {\bibinfo {title} {Theory of transport
  properties in the $p$-wave superconducting state of
  ${\mathrm{sr}}_{2}{\mathrm{ruo}}_{4}$ –a microscopic determination of the
  gap structure–},\ }\href@noop {} {\bibfield  {journal} {\bibinfo  {journal}
  {J. Phys. Soc. Jpn.}\ }\textbf {\bibinfo {volume} {74}},\ \bibinfo {pages}
  {1818} (\bibinfo {year} {2005})}\BibitemShut {NoStop}%
\bibitem [{\citenamefont {Nomura}\ \emph {et~al.}(2008)\citenamefont {Nomura},
  \citenamefont {Hirashima},\ and\ \citenamefont {Yamada}}]{NomuraJPSJ2008}%
  \BibitemOpen
  \bibfield  {author} {\bibinfo {author} {\bibfnamefont {T.}~\bibnamefont
  {Nomura}}, \bibinfo {author} {\bibfnamefont {D.~S.}\ \bibnamefont
  {Hirashima}},\ and\ \bibinfo {author} {\bibfnamefont {K.}~\bibnamefont
  {Yamada}},\ }\bibfield  {title} {\bibinfo {title} {Possible collective spin
  excitation in the spin-triplet superconducting state of
  ${\mathrm{sr}}_{2}{\mathrm{ruo}}_{4}$: Multi-band theory},\ }\href@noop {}
  {\bibfield  {journal} {\bibinfo  {journal} {J. Phys. Soc. Jpn.}\ }\textbf
  {\bibinfo {volume} {77}},\ \bibinfo {pages} {024701} (\bibinfo {year}
  {2008})}\BibitemShut {NoStop}%
\bibitem [{\citenamefont {Raghu}\ \emph {et~al.}(2010)\citenamefont {Raghu},
  \citenamefont {Kapitulnik},\ and\ \citenamefont {Kivelson}}]{RaghuPRL2010}%
  \BibitemOpen
  \bibfield  {author} {\bibinfo {author} {\bibfnamefont {S.}~\bibnamefont
  {Raghu}}, \bibinfo {author} {\bibfnamefont {A.}~\bibnamefont {Kapitulnik}},\
  and\ \bibinfo {author} {\bibfnamefont {S.~A.}\ \bibnamefont {Kivelson}},\
  }\bibfield  {title} {\bibinfo {title} {Hidden quasi-one-dimensional
  superconductivity in ${\mathrm{sr}}_{2}{\mathrm{ruo}}_{4}$},\ }\href
  {https://doi.org/10.1103/PhysRevLett.105.136401} {\bibfield  {journal}
  {\bibinfo  {journal} {Phys. Rev. Lett.}\ }\textbf {\bibinfo {volume} {105}},\
  \bibinfo {pages} {136401} (\bibinfo {year} {2010})}\BibitemShut {NoStop}%
\bibitem [{\citenamefont {Tsuchiizu}\ \emph {et~al.}(2015)\citenamefont
  {Tsuchiizu}, \citenamefont {Yamakawa}, \citenamefont {Onari}, \citenamefont
  {Ohno},\ and\ \citenamefont {Kontani}}]{TsuchiizuPRB2015}%
  \BibitemOpen
  \bibfield  {author} {\bibinfo {author} {\bibfnamefont {M.}~\bibnamefont
  {Tsuchiizu}}, \bibinfo {author} {\bibfnamefont {Y.}~\bibnamefont {Yamakawa}},
  \bibinfo {author} {\bibfnamefont {S.}~\bibnamefont {Onari}}, \bibinfo
  {author} {\bibfnamefont {Y.}~\bibnamefont {Ohno}},\ and\ \bibinfo {author}
  {\bibfnamefont {H.}~\bibnamefont {Kontani}},\ }\bibfield  {title} {\bibinfo
  {title} {Spin-triplet superconductivity in
  ${\mathrm{sr}}_{2}{\mathrm{ruo}}_{4}$ due to orbital and spin fluctuations:
  Analyses by two-dimensional renormalization group theory and self-consistent
  vertex-correction method},\ }\href
  {https://doi.org/10.1103/PhysRevB.91.155103} {\bibfield  {journal} {\bibinfo
  {journal} {Phys. Rev. B}\ }\textbf {\bibinfo {volume} {91}},\ \bibinfo
  {pages} {155103} (\bibinfo {year} {2015})}\BibitemShut {NoStop}%
\bibitem [{\citenamefont {Zhang}\ \emph {et~al.}(2018)\citenamefont {Zhang},
  \citenamefont {Huang}, \citenamefont {Yang},\ and\ \citenamefont
  {Yao}}]{ZhangPRB2018}%
  \BibitemOpen
  \bibfield  {author} {\bibinfo {author} {\bibfnamefont {L.-D.}\ \bibnamefont
  {Zhang}}, \bibinfo {author} {\bibfnamefont {W.}~\bibnamefont {Huang}},
  \bibinfo {author} {\bibfnamefont {F.}~\bibnamefont {Yang}},\ and\ \bibinfo
  {author} {\bibfnamefont {H.}~\bibnamefont {Yao}},\ }\bibfield  {title}
  {\bibinfo {title} {Superconducting pairing in
  ${\mathrm{sr}}_{2}{\mathrm{ruo}}_{4}$ from weak to intermediate coupling},\
  }\href {https://doi.org/10.1103/PhysRevB.97.060510} {\bibfield  {journal}
  {\bibinfo  {journal} {Phys. Rev. B}\ }\textbf {\bibinfo {volume} {97}},\
  \bibinfo {pages} {060510} (\bibinfo {year} {2018})}\BibitemShut {NoStop}%
\bibitem [{\citenamefont {Wang}\ \emph {et~al.}(2019)\citenamefont {Wang},
  \citenamefont {Zhang}, \citenamefont {Zhang},\ and\ \citenamefont
  {Wang}}]{WangPRL2019}%
  \BibitemOpen
  \bibfield  {author} {\bibinfo {author} {\bibfnamefont {W.-S.}\ \bibnamefont
  {Wang}}, \bibinfo {author} {\bibfnamefont {C.-C.}\ \bibnamefont {Zhang}},
  \bibinfo {author} {\bibfnamefont {F.-C.}\ \bibnamefont {Zhang}},\ and\
  \bibinfo {author} {\bibfnamefont {Q.-H.}\ \bibnamefont {Wang}},\ }\bibfield
  {title} {\bibinfo {title} {Theory of chiral $p$-wave superconductivity with
  near nodes for ${\mathrm{sr}}_{2}{\mathrm{ruo}}_{4}$},\ }\href
  {https://doi.org/10.1103/PhysRevLett.122.027002} {\bibfield  {journal}
  {\bibinfo  {journal} {Phys. Rev. Lett.}\ }\textbf {\bibinfo {volume} {122}},\
  \bibinfo {pages} {027002} (\bibinfo {year} {2019})}\BibitemShut {NoStop}%
\bibitem [{\citenamefont {Wang}\ \emph {et~al.}(2020)\citenamefont {Wang},
  \citenamefont {Wang},\ and\ \citenamefont {Kallin}}]{WangPRB2020}%
  \BibitemOpen
  \bibfield  {author} {\bibinfo {author} {\bibfnamefont {Z.}~\bibnamefont
  {Wang}}, \bibinfo {author} {\bibfnamefont {X.}~\bibnamefont {Wang}},\ and\
  \bibinfo {author} {\bibfnamefont {C.}~\bibnamefont {Kallin}},\ }\bibfield
  {title} {\bibinfo {title} {Spin-orbit coupling and spin-triplet pairing
  symmetry in ${\mathrm{sr}}_{2}{\mathrm{ruo}}_{4}$},\ }\href
  {https://doi.org/10.1103/PhysRevB.101.064507} {\bibfield  {journal} {\bibinfo
   {journal} {Phys. Rev. B}\ }\textbf {\bibinfo {volume} {101}},\ \bibinfo
  {pages} {064507} (\bibinfo {year} {2020})}\BibitemShut {NoStop}%
\bibitem [{\citenamefont {Pustogow}\ \emph {et~al.}(2019)\citenamefont
  {Pustogow}, \citenamefont {Luo}, \citenamefont {Chronister}, \citenamefont
  {Su}, \citenamefont {Sokolov}, \citenamefont {Jerzembeck}, \citenamefont
  {Mackenzie}, \citenamefont {Hicks}, \citenamefont {Kikugawa}, \citenamefont
  {Raghu}, \citenamefont {Bauer},\ and\ \citenamefont
  {Brown}}]{PustogowNature2019}%
  \BibitemOpen
  \bibfield  {author} {\bibinfo {author} {\bibfnamefont {A.}~\bibnamefont
  {Pustogow}}, \bibinfo {author} {\bibfnamefont {Y.}~\bibnamefont {Luo}},
  \bibinfo {author} {\bibfnamefont {A.}~\bibnamefont {Chronister}}, \bibinfo
  {author} {\bibfnamefont {Y.-S.}\ \bibnamefont {Su}}, \bibinfo {author}
  {\bibfnamefont {D.~A.}\ \bibnamefont {Sokolov}}, \bibinfo {author}
  {\bibfnamefont {F.}~\bibnamefont {Jerzembeck}}, \bibinfo {author}
  {\bibfnamefont {A.~P.}\ \bibnamefont {Mackenzie}}, \bibinfo {author}
  {\bibfnamefont {C.~W.}\ \bibnamefont {Hicks}}, \bibinfo {author}
  {\bibfnamefont {N.}~\bibnamefont {Kikugawa}}, \bibinfo {author}
  {\bibfnamefont {S.}~\bibnamefont {Raghu}}, \bibinfo {author} {\bibfnamefont
  {E.~D.}\ \bibnamefont {Bauer}},\ and\ \bibinfo {author} {\bibfnamefont
  {S.~E.}\ \bibnamefont {Brown}},\ }\bibfield  {title} {\bibinfo {title}
  {Constraints on the superconducting order parameter in
  ${\mathrm{sr}}_{2}{\mathrm{ruo}}_{4}$ from oxygen-17 nuclear magnetic
  resonance},\ }\href {https://doi.org/10.1038/s41586-019-1596-2} {\bibfield
  {journal} {\bibinfo  {journal} {Nature}\ }\textbf {\bibinfo {volume} {574}},\
  \bibinfo {pages} {72} (\bibinfo {year} {2019})}\BibitemShut {NoStop}%
\bibitem [{\citenamefont {Ishida}\ \emph {et~al.}(2020)\citenamefont {Ishida},
  \citenamefont {Manago}, \citenamefont {Kinjo},\ and\ \citenamefont
  {Maeno}}]{IshidaJPSJ2020}%
  \BibitemOpen
  \bibfield  {author} {\bibinfo {author} {\bibfnamefont {K.}~\bibnamefont
  {Ishida}}, \bibinfo {author} {\bibfnamefont {M.}~\bibnamefont {Manago}},
  \bibinfo {author} {\bibfnamefont {K.}~\bibnamefont {Kinjo}},\ and\ \bibinfo
  {author} {\bibfnamefont {Y.}~\bibnamefont {Maeno}},\ }\bibfield  {title}
  {\bibinfo {title} {Reduction of the 17o knight shift in the superconducting
  state and the heat-up effect by nmr pulses on
  ${\mathrm{sr}}_{2}{\mathrm{ruo}}_{4}$},\ }\href
  {https://doi.org/10.7566/JPSJ.89.034712} {\bibfield  {journal} {\bibinfo
  {journal} {J. Phys. Soc. Jpn.}\ }\textbf {\bibinfo {volume} {89}},\ \bibinfo
  {pages} {034712} (\bibinfo {year} {2020})}\BibitemShut {NoStop}%
\bibitem [{\citenamefont {Chronister}\ \emph {et~al.}(2021)\citenamefont
  {Chronister}, \citenamefont {Pustogow}, \citenamefont {Kikugawa},
  \citenamefont {Sokolov}, \citenamefont {Jerzembeck}, \citenamefont {Hicks},
  \citenamefont {Mackenzie}, \citenamefont {Bauer},\ and\ \citenamefont
  {Brown}}]{Chronistere2025313118}%
  \BibitemOpen
  \bibfield  {author} {\bibinfo {author} {\bibfnamefont {A.}~\bibnamefont
  {Chronister}}, \bibinfo {author} {\bibfnamefont {A.}~\bibnamefont
  {Pustogow}}, \bibinfo {author} {\bibfnamefont {N.}~\bibnamefont {Kikugawa}},
  \bibinfo {author} {\bibfnamefont {D.~A.}\ \bibnamefont {Sokolov}}, \bibinfo
  {author} {\bibfnamefont {F.}~\bibnamefont {Jerzembeck}}, \bibinfo {author}
  {\bibfnamefont {C.~W.}\ \bibnamefont {Hicks}}, \bibinfo {author}
  {\bibfnamefont {A.~P.}\ \bibnamefont {Mackenzie}}, \bibinfo {author}
  {\bibfnamefont {E.~D.}\ \bibnamefont {Bauer}},\ and\ \bibinfo {author}
  {\bibfnamefont {S.~E.}\ \bibnamefont {Brown}},\ }\bibfield  {title} {\bibinfo
  {title} {Evidence for even parity unconventional superconductivity in
  ${\mathrm{sr}}_{2}{\mathrm{ruo}}_{4}$},\ }\bibfield  {journal} {\bibinfo
  {journal} {Proceedings of the National Academy of Sciences}\ }\textbf
  {\bibinfo {volume} {118}},\ \href {https://doi.org/10.1073/pnas.2025313118}
  {10.1073/pnas.2025313118} (\bibinfo {year} {2021})\BibitemShut {NoStop}%
\bibitem [{\citenamefont {Leggett}\ and\ \citenamefont
  {Liu}(2021)}]{Leggett2021}%
  \BibitemOpen
  \bibfield  {author} {\bibinfo {author} {\bibfnamefont {A.~J.}\ \bibnamefont
  {Leggett}}\ and\ \bibinfo {author} {\bibfnamefont {Y.}~\bibnamefont {Liu}},\
  }\bibfield  {title} {\bibinfo {title} {Symmetry properties of superconducting
  order parameter in ${\mathrm{sr}}_{2}{\mathrm{ruo}}_{4}$},\ }\href
  {https://doi.org/10.1007/s10948-021-05865-3} {\bibfield  {journal} {\bibinfo
  {journal} {Journal of Superconductivity and Novel Magnetism}\ }\textbf
  {\bibinfo {volume} {34}},\ \bibinfo {pages} {1647} (\bibinfo {year}
  {2021})}\BibitemShut {NoStop}%
\bibitem [{\citenamefont {Ghosh}\ \emph {et~al.}(2021)\citenamefont {Ghosh},
  \citenamefont {Shekhter}, \citenamefont {Jerzembeck}, \citenamefont
  {Kikugawa}, \citenamefont {Sokolov}, \citenamefont {Brando}, \citenamefont
  {Mackenzie}, \citenamefont {Hicks},\ and\ \citenamefont
  {Ramshaw}}]{GhoshNatPhys2021}%
  \BibitemOpen
  \bibfield  {author} {\bibinfo {author} {\bibfnamefont {S.}~\bibnamefont
  {Ghosh}}, \bibinfo {author} {\bibfnamefont {A.}~\bibnamefont {Shekhter}},
  \bibinfo {author} {\bibfnamefont {F.}~\bibnamefont {Jerzembeck}}, \bibinfo
  {author} {\bibfnamefont {N.}~\bibnamefont {Kikugawa}}, \bibinfo {author}
  {\bibfnamefont {D.~A.}\ \bibnamefont {Sokolov}}, \bibinfo {author}
  {\bibfnamefont {M.}~\bibnamefont {Brando}}, \bibinfo {author} {\bibfnamefont
  {A.~P.}\ \bibnamefont {Mackenzie}}, \bibinfo {author} {\bibfnamefont {C.~W.}\
  \bibnamefont {Hicks}},\ and\ \bibinfo {author} {\bibfnamefont {B.~J.}\
  \bibnamefont {Ramshaw}},\ }\bibfield  {title} {\bibinfo {title}
  {Thermodynamic evidence for a two-component superconducting order parameter
  in ${\mathrm{sr}}_{2}{\mathrm{ruo}}_{4}$},\ }\href
  {https://doi.org/10.1038/s41567-020-1032-4} {\bibfield  {journal} {\bibinfo
  {journal} {Nat. Phys.}\ }\textbf {\bibinfo {volume} {17}},\ \bibinfo {pages}
  {199} (\bibinfo {year} {2021})}\BibitemShut {NoStop}%
\bibitem [{\citenamefont {Agterberg}(2021)}]{AgterbergNatPhysics2021}%
  \BibitemOpen
  \bibfield  {author} {\bibinfo {author} {\bibfnamefont {D.~F.}\ \bibnamefont
  {Agterberg}},\ }\bibfield  {title} {\bibinfo {title} {The symmetry of
  superconducting ${\mathrm{sr}}_{2}{\mathrm{ruo}}_{4}$},\ }\href@noop {}
  {\bibfield  {journal} {\bibinfo  {journal} {Nat. Phys.}\ }\textbf {\bibinfo
  {volume} {17}},\ \bibinfo {pages} {169} (\bibinfo {year} {2021})}\BibitemShut
  {NoStop}%
\bibitem [{\citenamefont {Benhabib}\ \emph {et~al.}(2021)\citenamefont
  {Benhabib}, \citenamefont {Lupien}, \citenamefont {Paul}, \citenamefont
  {Berges}, \citenamefont {Dion}, \citenamefont {Nardone}, \citenamefont
  {Zitouni}, \citenamefont {Mao}, \citenamefont {Maeno}, \citenamefont
  {Georges}, \citenamefont {Taillefer},\ and\ \citenamefont
  {Proust}}]{BenhabibNatPhys2021}%
  \BibitemOpen
  \bibfield  {author} {\bibinfo {author} {\bibfnamefont {S.}~\bibnamefont
  {Benhabib}}, \bibinfo {author} {\bibfnamefont {C.}~\bibnamefont {Lupien}},
  \bibinfo {author} {\bibfnamefont {I.}~\bibnamefont {Paul}}, \bibinfo {author}
  {\bibfnamefont {L.}~\bibnamefont {Berges}}, \bibinfo {author} {\bibfnamefont
  {M.}~\bibnamefont {Dion}}, \bibinfo {author} {\bibfnamefont {M.}~\bibnamefont
  {Nardone}}, \bibinfo {author} {\bibfnamefont {A.}~\bibnamefont {Zitouni}},
  \bibinfo {author} {\bibfnamefont {Z.~Q.}\ \bibnamefont {Mao}}, \bibinfo
  {author} {\bibfnamefont {Y.}~\bibnamefont {Maeno}}, \bibinfo {author}
  {\bibfnamefont {A.}~\bibnamefont {Georges}}, \bibinfo {author} {\bibfnamefont
  {L.}~\bibnamefont {Taillefer}},\ and\ \bibinfo {author} {\bibfnamefont
  {C.}~\bibnamefont {Proust}},\ }\bibfield  {title} {\bibinfo {title}
  {Ultrasound evidence for a two-component superconducting order parameter in
  ${\mathrm{sr}}_{2}{\mathrm{ruo}}_{4}$},\ }\href
  {https://doi.org/10.1038/s41567-020-1033-3} {\bibfield  {journal} {\bibinfo
  {journal} {Nat. Phys.}\ }\textbf {\bibinfo {volume} {17}},\ \bibinfo {pages}
  {194} (\bibinfo {year} {2021})}\BibitemShut {NoStop}%
\bibitem [{\citenamefont {R\o{}mer}\ \emph {et~al.}(2019)\citenamefont
  {R\o{}mer}, \citenamefont {Scherer}, \citenamefont {Eremin}, \citenamefont
  {Hirschfeld},\ and\ \citenamefont {Andersen}}]{RomerPRL2019}%
  \BibitemOpen
  \bibfield  {author} {\bibinfo {author} {\bibfnamefont {A.~T.}\ \bibnamefont
  {R\o{}mer}}, \bibinfo {author} {\bibfnamefont {D.~D.}\ \bibnamefont
  {Scherer}}, \bibinfo {author} {\bibfnamefont {I.~M.}\ \bibnamefont {Eremin}},
  \bibinfo {author} {\bibfnamefont {P.~J.}\ \bibnamefont {Hirschfeld}},\ and\
  \bibinfo {author} {\bibfnamefont {B.~M.}\ \bibnamefont {Andersen}},\
  }\bibfield  {title} {\bibinfo {title} {Knight shift and leading
  superconducting instability from spin fluctuations in
  ${\mathrm{sr}}_{2}{\mathrm{ruo}}_{4}$},\ }\href
  {https://doi.org/10.1103/PhysRevLett.123.247001} {\bibfield  {journal}
  {\bibinfo  {journal} {Phys. Rev. Lett.}\ }\textbf {\bibinfo {volume} {123}},\
  \bibinfo {pages} {247001} (\bibinfo {year} {2019})}\BibitemShut {NoStop}%
\bibitem [{\citenamefont {Kivelson}\ \emph {et~al.}(2020)\citenamefont
  {Kivelson}, \citenamefont {Yuan}, \citenamefont {Ramshaw},\ and\
  \citenamefont {Thomale}}]{KivelsonNPJ2020}%
  \BibitemOpen
  \bibfield  {author} {\bibinfo {author} {\bibfnamefont {S.~A.}\ \bibnamefont
  {Kivelson}}, \bibinfo {author} {\bibfnamefont {A.~C.}\ \bibnamefont {Yuan}},
  \bibinfo {author} {\bibfnamefont {B.}~\bibnamefont {Ramshaw}},\ and\ \bibinfo
  {author} {\bibfnamefont {R.}~\bibnamefont {Thomale}},\ }\bibfield  {title}
  {\bibinfo {title} {A proposal for reconciling diverse experiments on the
  superconducting state in ${\mathrm{sr}}_{2}{\mathrm{ruo}}_{4}$},\ }\href
  {https://doi.org/10.1038/s41535-020-0245-1} {\bibfield  {journal} {\bibinfo
  {journal} {npj Quantum Materials}\ }\textbf {\bibinfo {volume} {5}},\
  \bibinfo {pages} {43} (\bibinfo {year} {2020})}\BibitemShut {NoStop}%
\bibitem [{\citenamefont {Willa}\ \emph {et~al.}(2021)\citenamefont {Willa},
  \citenamefont {Hecker}, \citenamefont {Fernandes},\ and\ \citenamefont
  {Schmalian}}]{WillaPRB2021}%
  \BibitemOpen
  \bibfield  {author} {\bibinfo {author} {\bibfnamefont {R.}~\bibnamefont
  {Willa}}, \bibinfo {author} {\bibfnamefont {M.}~\bibnamefont {Hecker}},
  \bibinfo {author} {\bibfnamefont {R.~M.}\ \bibnamefont {Fernandes}},\ and\
  \bibinfo {author} {\bibfnamefont {J.}~\bibnamefont {Schmalian}},\ }\bibfield
  {title} {\bibinfo {title} {Inhomogeneous time-reversal symmetry breaking in
  ${\mathrm{sr}}_{2}{\mathrm{ruo}}_{4}$},\ }\href
  {https://doi.org/10.1103/PhysRevB.104.024511} {\bibfield  {journal} {\bibinfo
   {journal} {Phys. Rev. B}\ }\textbf {\bibinfo {volume} {104}},\ \bibinfo
  {pages} {024511} (\bibinfo {year} {2021})}\BibitemShut {NoStop}%
\bibitem [{\citenamefont {Clepkens}\ \emph
  {et~al.}(2021{\natexlab{a}})\citenamefont {Clepkens}, \citenamefont
  {Lindquist}, \citenamefont {Liu},\ and\ \citenamefont
  {Kee}}]{clepkens2021higher}%
  \BibitemOpen
  \bibfield  {author} {\bibinfo {author} {\bibfnamefont {J.}~\bibnamefont
  {Clepkens}}, \bibinfo {author} {\bibfnamefont {A.~W.}\ \bibnamefont
  {Lindquist}}, \bibinfo {author} {\bibfnamefont {X.}~\bibnamefont {Liu}},\
  and\ \bibinfo {author} {\bibfnamefont {H.-Y.}\ \bibnamefont {Kee}},\
  }\href@noop {} {\bibinfo {title} {Higher angular momentum pairings in
  inter-orbital shadowed-triplet superconductors: Application to
  sr$_{2}$ruo$_{4}$}} (\bibinfo {year} {2021}{\natexlab{a}}),\ \Eprint
  {https://arxiv.org/abs/2107.00047} {arXiv:2107.00047 [cond-mat.supr-con]}
  \BibitemShut {NoStop}%
\bibitem [{\citenamefont {Yuan}\ \emph {et~al.}(2021)\citenamefont {Yuan},
  \citenamefont {Berg},\ and\ \citenamefont {Kivelson}}]{YuanPRB2021}%
  \BibitemOpen
  \bibfield  {author} {\bibinfo {author} {\bibfnamefont {A.~C.}\ \bibnamefont
  {Yuan}}, \bibinfo {author} {\bibfnamefont {E.}~\bibnamefont {Berg}},\ and\
  \bibinfo {author} {\bibfnamefont {S.~A.}\ \bibnamefont {Kivelson}},\
  }\bibfield  {title} {\bibinfo {title} {Strain-induced time reversal breaking
  and half quantum vortices near a putative superconducting tetracritical point
  in ${\mathrm{sr}}_{2}{\mathrm{ruo}}_{4}$},\ }\href
  {https://doi.org/10.1103/PhysRevB.104.054518} {\bibfield  {journal} {\bibinfo
   {journal} {Phys. Rev. B}\ }\textbf {\bibinfo {volume} {104}},\ \bibinfo
  {pages} {054518} (\bibinfo {year} {2021})}\BibitemShut {NoStop}%
\bibitem [{\citenamefont {Clepkens}\ \emph
  {et~al.}(2021{\natexlab{b}})\citenamefont {Clepkens}, \citenamefont
  {Lindquist},\ and\ \citenamefont {Kee}}]{ClepkensPRR2021}%
  \BibitemOpen
  \bibfield  {author} {\bibinfo {author} {\bibfnamefont {J.}~\bibnamefont
  {Clepkens}}, \bibinfo {author} {\bibfnamefont {A.~W.}\ \bibnamefont
  {Lindquist}},\ and\ \bibinfo {author} {\bibfnamefont {H.-Y.}\ \bibnamefont
  {Kee}},\ }\bibfield  {title} {\bibinfo {title} {Shadowed triplet pairings in
  hund's metals with spin-orbit coupling},\ }\href
  {https://doi.org/10.1103/PhysRevResearch.3.013001} {\bibfield  {journal}
  {\bibinfo  {journal} {Phys. Rev. Research}\ }\textbf {\bibinfo {volume}
  {3}},\ \bibinfo {pages} {013001} (\bibinfo {year}
  {2021}{\natexlab{b}})}\BibitemShut {NoStop}%
\bibitem [{\citenamefont {R\o{}mer}\ \emph {et~al.}(2021)\citenamefont
  {R\o{}mer}, \citenamefont {Hirschfeld},\ and\ \citenamefont
  {Andersen}}]{RomerPRB2021}%
  \BibitemOpen
  \bibfield  {author} {\bibinfo {author} {\bibfnamefont {A.~T.}\ \bibnamefont
  {R\o{}mer}}, \bibinfo {author} {\bibfnamefont {P.~J.}\ \bibnamefont
  {Hirschfeld}},\ and\ \bibinfo {author} {\bibfnamefont {B.~M.}\ \bibnamefont
  {Andersen}},\ }\bibfield  {title} {\bibinfo {title} {Superconducting state of
  ${\mathrm{sr}}_{2}{\mathrm{ruo}}_{4}$ in the presence of longer-range coulomb
  interactions},\ }\href {https://doi.org/10.1103/PhysRevB.104.064507}
  {\bibfield  {journal} {\bibinfo  {journal} {Phys. Rev. B}\ }\textbf {\bibinfo
  {volume} {104}},\ \bibinfo {pages} {064507} (\bibinfo {year}
  {2021})}\BibitemShut {NoStop}%
\bibitem [{\citenamefont {Suh}\ \emph {et~al.}(2020)\citenamefont {Suh},
  \citenamefont {Menke}, \citenamefont {Brydon}, \citenamefont {Timm},
  \citenamefont {Ramires},\ and\ \citenamefont {Agterberg}}]{SuhPRR2020}%
  \BibitemOpen
  \bibfield  {author} {\bibinfo {author} {\bibfnamefont {H.~G.}\ \bibnamefont
  {Suh}}, \bibinfo {author} {\bibfnamefont {H.}~\bibnamefont {Menke}}, \bibinfo
  {author} {\bibfnamefont {P.~M.~R.}\ \bibnamefont {Brydon}}, \bibinfo {author}
  {\bibfnamefont {C.}~\bibnamefont {Timm}}, \bibinfo {author} {\bibfnamefont
  {A.}~\bibnamefont {Ramires}},\ and\ \bibinfo {author} {\bibfnamefont {D.~F.}\
  \bibnamefont {Agterberg}},\ }\bibfield  {title} {\bibinfo {title}
  {Stabilizing even-parity chiral superconductivity in
  ${\mathrm{sr}}_{2}{\mathrm{ruo}}_{4}$},\ }\href
  {https://doi.org/10.1103/PhysRevResearch.2.032023} {\bibfield  {journal}
  {\bibinfo  {journal} {Phys. Rev. Research}\ }\textbf {\bibinfo {volume}
  {2}},\ \bibinfo {pages} {032023} (\bibinfo {year} {2020})}\BibitemShut
  {NoStop}%
\bibitem [{\citenamefont {Agterberg}\ \emph {et~al.}(2017)\citenamefont
  {Agterberg}, \citenamefont {Brydon},\ and\ \citenamefont
  {Timm}}]{AgterbergPRL2017}%
  \BibitemOpen
  \bibfield  {author} {\bibinfo {author} {\bibfnamefont {D.~F.}\ \bibnamefont
  {Agterberg}}, \bibinfo {author} {\bibfnamefont {P.~M.~R.}\ \bibnamefont
  {Brydon}},\ and\ \bibinfo {author} {\bibfnamefont {C.}~\bibnamefont {Timm}},\
  }\bibfield  {title} {\bibinfo {title} {Bogoliubov fermi surfaces in
  superconductors with broken time-reversal symmetry},\ }\href
  {https://doi.org/10.1103/PhysRevLett.118.127001} {\bibfield  {journal}
  {\bibinfo  {journal} {Phys. Rev. Lett.}\ }\textbf {\bibinfo {volume} {118}},\
  \bibinfo {pages} {127001} (\bibinfo {year} {2017})}\BibitemShut {NoStop}%
\bibitem [{\citenamefont {Brydon}\ \emph {et~al.}(2018)\citenamefont {Brydon},
  \citenamefont {Agterberg}, \citenamefont {Menke},\ and\ \citenamefont
  {Timm}}]{BrydonPRB2018}%
  \BibitemOpen
  \bibfield  {author} {\bibinfo {author} {\bibfnamefont {P.~M.~R.}\
  \bibnamefont {Brydon}}, \bibinfo {author} {\bibfnamefont {D.~F.}\
  \bibnamefont {Agterberg}}, \bibinfo {author} {\bibfnamefont {H.}~\bibnamefont
  {Menke}},\ and\ \bibinfo {author} {\bibfnamefont {C.}~\bibnamefont {Timm}},\
  }\bibfield  {title} {\bibinfo {title} {Bogoliubov fermi surfaces: General
  theory, magnetic order, and topology},\ }\href
  {https://doi.org/10.1103/PhysRevB.98.224509} {\bibfield  {journal} {\bibinfo
  {journal} {Phys. Rev. B}\ }\textbf {\bibinfo {volume} {98}},\ \bibinfo
  {pages} {224509} (\bibinfo {year} {2018})}\BibitemShut {NoStop}%
\bibitem [{\citenamefont {Puetter}\ and\ \citenamefont
  {Kee}(2012)}]{PuetterEPL2012}%
  \BibitemOpen
  \bibfield  {author} {\bibinfo {author} {\bibfnamefont {C.~M.}\ \bibnamefont
  {Puetter}}\ and\ \bibinfo {author} {\bibfnamefont {H.-Y.}\ \bibnamefont
  {Kee}},\ }\bibfield  {title} {\bibinfo {title} {Identifying spin-triplet
  pairing in spin-orbit coupled multi-band superconductors},\ }\href
  {https://doi.org/10.1209/0295-5075/98/27010} {\bibfield  {journal} {\bibinfo
  {journal} {{EPL} (Europhysics Letters)}\ }\textbf {\bibinfo {volume} {98}},\
  \bibinfo {pages} {27010} (\bibinfo {year} {2012})}\BibitemShut {NoStop}%
\bibitem [{\citenamefont {Ramires}\ and\ \citenamefont
  {Sigrist}(2019)}]{RamiresPRB2019}%
  \BibitemOpen
  \bibfield  {author} {\bibinfo {author} {\bibfnamefont {A.}~\bibnamefont
  {Ramires}}\ and\ \bibinfo {author} {\bibfnamefont {M.}~\bibnamefont
  {Sigrist}},\ }\bibfield  {title} {\bibinfo {title} {Superconducting order
  parameter of ${\mathrm{sr}}_{2}{\mathrm{ruo}}_{4}$: A microscopic
  perspective},\ }\href {https://doi.org/10.1103/PhysRevB.100.104501}
  {\bibfield  {journal} {\bibinfo  {journal} {Phys. Rev. B}\ }\textbf {\bibinfo
  {volume} {100}},\ \bibinfo {pages} {104501} (\bibinfo {year}
  {2019})}\BibitemShut {NoStop}%
\bibitem [{\citenamefont {Chen}\ and\ \citenamefont {An}(2020)}]{ChenPRB2020}%
  \BibitemOpen
  \bibfield  {author} {\bibinfo {author} {\bibfnamefont {W.}~\bibnamefont
  {Chen}}\ and\ \bibinfo {author} {\bibfnamefont {J.}~\bibnamefont {An}},\
  }\bibfield  {title} {\bibinfo {title} {Interorbital $p$- and $d$-wave
  pairings between ${d}_{xz/yz}$ and ${d}_{xy}$ orbitals in
  ${\mathrm{sr}}_{2}{\mathrm{ruo}}_{4}$},\ }\href
  {https://doi.org/10.1103/PhysRevB.102.094501} {\bibfield  {journal} {\bibinfo
   {journal} {Phys. Rev. B}\ }\textbf {\bibinfo {volume} {102}},\ \bibinfo
  {pages} {094501} (\bibinfo {year} {2020})}\BibitemShut {NoStop}%
\bibitem [{\citenamefont {Grinenko}\ \emph {et~al.}(2021)\citenamefont
  {Grinenko}, \citenamefont {Das}, \citenamefont {Gupta}, \citenamefont
  {Zinkl}, \citenamefont {Kikugawa}, \citenamefont {Maeno}, \citenamefont
  {Hicks}, \citenamefont {Klauss}, \citenamefont {Sigrist},\ and\ \citenamefont
  {Khasanov}}]{GrinenkoNatComm2021}%
  \BibitemOpen
  \bibfield  {author} {\bibinfo {author} {\bibfnamefont {V.}~\bibnamefont
  {Grinenko}}, \bibinfo {author} {\bibfnamefont {D.}~\bibnamefont {Das}},
  \bibinfo {author} {\bibfnamefont {R.}~\bibnamefont {Gupta}}, \bibinfo
  {author} {\bibfnamefont {B.}~\bibnamefont {Zinkl}}, \bibinfo {author}
  {\bibfnamefont {N.}~\bibnamefont {Kikugawa}}, \bibinfo {author}
  {\bibfnamefont {Y.}~\bibnamefont {Maeno}}, \bibinfo {author} {\bibfnamefont
  {C.~W.}\ \bibnamefont {Hicks}}, \bibinfo {author} {\bibfnamefont {H.-H.}\
  \bibnamefont {Klauss}}, \bibinfo {author} {\bibfnamefont {M.}~\bibnamefont
  {Sigrist}},\ and\ \bibinfo {author} {\bibfnamefont {R.}~\bibnamefont
  {Khasanov}},\ }\bibfield  {title} {\bibinfo {title} {Unsplit superconducting
  and time reversal symmetry breaking transitions in
  ${\text{sr}}_{2}{\text{ruo}}_{4}$ under hydrostatic pressure and disorder},\
  }\href@noop {} {\bibfield  {journal} {\bibinfo  {journal} {Nat. Commun.}\
  }\textbf {\bibinfo {volume} {12}},\ \bibinfo {pages} {3920} (\bibinfo {year}
  {2021})}\BibitemShut {NoStop}%
\bibitem [{\citenamefont {Yu}\ \emph {et~al.}(2018)\citenamefont {Yu},
  \citenamefont {Cheung}, \citenamefont {Raghu},\ and\ \citenamefont
  {Agterberg}}]{YuPRB2018}%
  \BibitemOpen
  \bibfield  {author} {\bibinfo {author} {\bibfnamefont {Y.}~\bibnamefont
  {Yu}}, \bibinfo {author} {\bibfnamefont {A.~K.~C.}\ \bibnamefont {Cheung}},
  \bibinfo {author} {\bibfnamefont {S.}~\bibnamefont {Raghu}},\ and\ \bibinfo
  {author} {\bibfnamefont {D.~F.}\ \bibnamefont {Agterberg}},\ }\bibfield
  {title} {\bibinfo {title} {Residual spin susceptibility in the spin-triplet
  orbital-singlet model},\ }\href {https://doi.org/10.1103/PhysRevB.98.184507}
  {\bibfield  {journal} {\bibinfo  {journal} {Phys. Rev. B}\ }\textbf {\bibinfo
  {volume} {98}},\ \bibinfo {pages} {184507} (\bibinfo {year}
  {2018})}\BibitemShut {NoStop}%
\bibitem [{\citenamefont {Lindquist}\ and\ \citenamefont
  {Kee}(2020)}]{LindquistPRR2020}%
  \BibitemOpen
  \bibfield  {author} {\bibinfo {author} {\bibfnamefont {A.~W.}\ \bibnamefont
  {Lindquist}}\ and\ \bibinfo {author} {\bibfnamefont {H.-Y.}\ \bibnamefont
  {Kee}},\ }\bibfield  {title} {\bibinfo {title} {Distinct reduction of knight
  shift in superconducting state of ${\mathrm{sr}}_{2}{\mathrm{ruo}}_{4}$ under
  uniaxial strain},\ }\href {https://doi.org/10.1103/PhysRevResearch.2.032055}
  {\bibfield  {journal} {\bibinfo  {journal} {Phys. Rev. Research}\ }\textbf
  {\bibinfo {volume} {2}},\ \bibinfo {pages} {032055} (\bibinfo {year}
  {2020})}\BibitemShut {NoStop}%
\bibitem [{\citenamefont {Scaffidi}\ \emph {et~al.}(2014)\citenamefont
  {Scaffidi}, \citenamefont {Romers},\ and\ \citenamefont
  {Simon}}]{ScaffidiPRB2014}%
  \BibitemOpen
  \bibfield  {author} {\bibinfo {author} {\bibfnamefont {T.}~\bibnamefont
  {Scaffidi}}, \bibinfo {author} {\bibfnamefont {J.~C.}\ \bibnamefont
  {Romers}},\ and\ \bibinfo {author} {\bibfnamefont {S.~H.}\ \bibnamefont
  {Simon}},\ }\bibfield  {title} {\bibinfo {title} {Pairing symmetry and
  dominant band in ${\mathrm{sr}}_{2}{\mathrm{ruo}}_{4}$},\ }\href
  {https://doi.org/10.1103/PhysRevB.89.220510} {\bibfield  {journal} {\bibinfo
  {journal} {Phys. Rev. B}\ }\textbf {\bibinfo {volume} {89}},\ \bibinfo
  {pages} {220510} (\bibinfo {year} {2014})}\BibitemShut {NoStop}%
\bibitem [{\citenamefont {Ramires}\ and\ \citenamefont
  {Sigrist}(2016)}]{RamiresPRB2016}%
  \BibitemOpen
  \bibfield  {author} {\bibinfo {author} {\bibfnamefont {A.}~\bibnamefont
  {Ramires}}\ and\ \bibinfo {author} {\bibfnamefont {M.}~\bibnamefont
  {Sigrist}},\ }\bibfield  {title} {\bibinfo {title} {Identifying detrimental
  effects for multiorbital superconductivity: Application to
  ${\mathrm{sr}}_{2}{\mathrm{ruo}}_{4}$},\ }\href
  {https://doi.org/10.1103/PhysRevB.94.104501} {\bibfield  {journal} {\bibinfo
  {journal} {Phys. Rev. B}\ }\textbf {\bibinfo {volume} {94}},\ \bibinfo
  {pages} {104501} (\bibinfo {year} {2016})}\BibitemShut {NoStop}%
\bibitem [{\citenamefont {Haverkort}\ \emph {et~al.}(2008)\citenamefont
  {Haverkort}, \citenamefont {Elfimov}, \citenamefont {Tjeng}, \citenamefont
  {Sawatzky},\ and\ \citenamefont {Damascelli}}]{HaverkortPRL2008}%
  \BibitemOpen
  \bibfield  {author} {\bibinfo {author} {\bibfnamefont {M.~W.}\ \bibnamefont
  {Haverkort}}, \bibinfo {author} {\bibfnamefont {I.~S.}\ \bibnamefont
  {Elfimov}}, \bibinfo {author} {\bibfnamefont {L.~H.}\ \bibnamefont {Tjeng}},
  \bibinfo {author} {\bibfnamefont {G.~A.}\ \bibnamefont {Sawatzky}},\ and\
  \bibinfo {author} {\bibfnamefont {A.}~\bibnamefont {Damascelli}},\ }\bibfield
   {title} {\bibinfo {title} {Strong spin-orbit coupling effects on the fermi
  surface of ${\mathrm{sr}}_{2}{\mathrm{ruo}}_{4}$ and
  ${\mathrm{sr}}_{2}{\mathrm{rho}}_{4}$},\ }\href
  {https://doi.org/10.1103/PhysRevLett.101.026406} {\bibfield  {journal}
  {\bibinfo  {journal} {Phys. Rev. Lett.}\ }\textbf {\bibinfo {volume} {101}},\
  \bibinfo {pages} {026406} (\bibinfo {year} {2008})}\BibitemShut {NoStop}%
\bibitem [{\citenamefont {Veenstra}\ \emph {et~al.}(2014)\citenamefont
  {Veenstra}, \citenamefont {Zhu}, \citenamefont {Raichle}, \citenamefont
  {Ludbrook}, \citenamefont {Nicolaou}, \citenamefont {Slomski}, \citenamefont
  {Landolt}, \citenamefont {Kittaka}, \citenamefont {Maeno}, \citenamefont
  {Dil}, \citenamefont {Elfimov}, \citenamefont {Haverkort},\ and\
  \citenamefont {Damascelli}}]{VeenstraPRL2014}%
  \BibitemOpen
  \bibfield  {author} {\bibinfo {author} {\bibfnamefont {C.~N.}\ \bibnamefont
  {Veenstra}}, \bibinfo {author} {\bibfnamefont {Z.-H.}\ \bibnamefont {Zhu}},
  \bibinfo {author} {\bibfnamefont {M.}~\bibnamefont {Raichle}}, \bibinfo
  {author} {\bibfnamefont {B.~M.}\ \bibnamefont {Ludbrook}}, \bibinfo {author}
  {\bibfnamefont {A.}~\bibnamefont {Nicolaou}}, \bibinfo {author}
  {\bibfnamefont {B.}~\bibnamefont {Slomski}}, \bibinfo {author} {\bibfnamefont
  {G.}~\bibnamefont {Landolt}}, \bibinfo {author} {\bibfnamefont
  {S.}~\bibnamefont {Kittaka}}, \bibinfo {author} {\bibfnamefont
  {Y.}~\bibnamefont {Maeno}}, \bibinfo {author} {\bibfnamefont {J.~H.}\
  \bibnamefont {Dil}}, \bibinfo {author} {\bibfnamefont {I.~S.}\ \bibnamefont
  {Elfimov}}, \bibinfo {author} {\bibfnamefont {M.~W.}\ \bibnamefont
  {Haverkort}},\ and\ \bibinfo {author} {\bibfnamefont {A.}~\bibnamefont
  {Damascelli}},\ }\bibfield  {title} {\bibinfo {title} {Spin-orbital
  entanglement and the breakdown of singlets and triplets in
  ${\mathrm{sr}}_{2}{\mathrm{ruo}}_{4}$ revealed by spin- and angle-resolved
  photoemission spectroscopy},\ }\href
  {https://doi.org/10.1103/PhysRevLett.112.127002} {\bibfield  {journal}
  {\bibinfo  {journal} {Phys. Rev. Lett.}\ }\textbf {\bibinfo {volume} {112}},\
  \bibinfo {pages} {127002} (\bibinfo {year} {2014})}\BibitemShut {NoStop}%
\bibitem [{\citenamefont {R\o{}ising}\ \emph {et~al.}(2019)\citenamefont
  {R\o{}ising}, \citenamefont {Scaffidi}, \citenamefont {Flicker},
  \citenamefont {Lange},\ and\ \citenamefont {Simon}}]{RosisingPRR2019}%
  \BibitemOpen
  \bibfield  {author} {\bibinfo {author} {\bibfnamefont {H.~S.}\ \bibnamefont
  {R\o{}ising}}, \bibinfo {author} {\bibfnamefont {T.}~\bibnamefont
  {Scaffidi}}, \bibinfo {author} {\bibfnamefont {F.}~\bibnamefont {Flicker}},
  \bibinfo {author} {\bibfnamefont {G.~F.}\ \bibnamefont {Lange}},\ and\
  \bibinfo {author} {\bibfnamefont {S.~H.}\ \bibnamefont {Simon}},\ }\bibfield
  {title} {\bibinfo {title} {Superconducting order of
  ${\mathrm{sr}}_{2}{\mathrm{ruo}}_{4}$ from a three-dimensional microscopic
  model},\ }\href {https://doi.org/10.1103/PhysRevResearch.1.033108} {\bibfield
   {journal} {\bibinfo  {journal} {Phys. Rev. Research}\ }\textbf {\bibinfo
  {volume} {1}},\ \bibinfo {pages} {033108} (\bibinfo {year}
  {2019})}\BibitemShut {NoStop}%
\bibitem [{\citenamefont {Berezinskii}(1974)}]{Berezinskii}%
  \BibitemOpen
  \bibfield  {author} {\bibinfo {author} {\bibfnamefont {V.~L.}\ \bibnamefont
  {Berezinskii}},\ }\bibfield  {title} {\bibinfo {title} {New model of the
  anisotropy phase of superfluid he3},\ }\href@noop {} {\bibfield  {journal}
  {\bibinfo  {journal} {JETP Lett.}\ }\textbf {\bibinfo {volume} {20}},\
  \bibinfo {pages} {287} (\bibinfo {year} {1974})}\BibitemShut {NoStop}%
\bibitem [{\citenamefont {Balatsky}\ and\ \citenamefont
  {Abrahams}(1992)}]{Balatsky}%
  \BibitemOpen
  \bibfield  {author} {\bibinfo {author} {\bibfnamefont {A.}~\bibnamefont
  {Balatsky}}\ and\ \bibinfo {author} {\bibfnamefont {E.}~\bibnamefont
  {Abrahams}},\ }\bibfield  {title} {\bibinfo {title} {New class of singlet
  superconductors which break the time reversal and parity},\ }\href@noop {}
  {\bibfield  {journal} {\bibinfo  {journal} {Phys. Rev. B}\ }\textbf {\bibinfo
  {volume} {45}},\ \bibinfo {pages} {13125} (\bibinfo {year}
  {1992})}\BibitemShut {NoStop}%
\bibitem [{\citenamefont {Shigeta}\ \emph {et~al.}(2009)\citenamefont
  {Shigeta}, \citenamefont {Onari}, \citenamefont {Yada},\ and\ \citenamefont
  {Tanaka}}]{Shigeta}%
  \BibitemOpen
  \bibfield  {author} {\bibinfo {author} {\bibfnamefont {K.}~\bibnamefont
  {Shigeta}}, \bibinfo {author} {\bibfnamefont {S.}~\bibnamefont {Onari}},
  \bibinfo {author} {\bibfnamefont {K.}~\bibnamefont {Yada}},\ and\ \bibinfo
  {author} {\bibfnamefont {Y.}~\bibnamefont {Tanaka}},\ }\bibfield  {title}
  {\bibinfo {title} {Theory of odd-frequency pairings on a
  quasi-one-dimensional lattice in the hubbard model},\ }\href@noop {}
  {\bibfield  {journal} {\bibinfo  {journal} {Phys. Rev. B}\ }\textbf {\bibinfo
  {volume} {79}},\ \bibinfo {pages} {174507} (\bibinfo {year}
  {2009})}\BibitemShut {NoStop}%
\bibitem [{\citenamefont {Tanaka}\ \emph {et~al.}(2012)\citenamefont {Tanaka},
  \citenamefont {Sato},\ and\ \citenamefont {Nagaosa}}]{tanaka12}%
  \BibitemOpen
  \bibfield  {author} {\bibinfo {author} {\bibfnamefont {Y.}~\bibnamefont
  {Tanaka}}, \bibinfo {author} {\bibfnamefont {M.}~\bibnamefont {Sato}},\ and\
  \bibinfo {author} {\bibfnamefont {N.}~\bibnamefont {Nagaosa}},\ }\bibfield
  {title} {\bibinfo {title} {Symmetry and topology in superconductors
  –odd-frequency pairing and edge states–},\ }\href@noop {} {\bibfield
  {journal} {\bibinfo  {journal} {J. Phys. Soc. Jpn.}\ }\textbf {\bibinfo
  {volume} {81}},\ \bibinfo {pages} {011013} (\bibinfo {year}
  {2012})}\BibitemShut {NoStop}%
\bibitem [{\citenamefont {Linder}\ and\ \citenamefont
  {Balatsky}(2019)}]{LinderRMP2019}%
  \BibitemOpen
  \bibfield  {author} {\bibinfo {author} {\bibfnamefont {J.}~\bibnamefont
  {Linder}}\ and\ \bibinfo {author} {\bibfnamefont {A.}~\bibnamefont
  {Balatsky}},\ }\bibfield  {title} {\bibinfo {title} {Odd-frequency
  superconductivity},\ }\href {https://doi.org/10.1103/RevModPhys.91.045005}
  {\bibfield  {journal} {\bibinfo  {journal} {Rev. Mod. Phys.}\ }\textbf
  {\bibinfo {volume} {91}},\ \bibinfo {pages} {045005} (\bibinfo {year}
  {2019})}\BibitemShut {NoStop}%
\bibitem [{\citenamefont {S.~Hirashima}(2007)}]{HirashimaJPSJ2007}%
  \BibitemOpen
  \bibfield  {author} {\bibinfo {author} {\bibfnamefont {D.}~\bibnamefont
  {S.~Hirashima}},\ }\bibfield  {title} {\bibinfo {title} {Dynamical spin
  susceptibilities in the superconducting phase of
  ${\mathrm{sr}}_{2}{\mathrm{ruo}}_{4}$},\ }\href
  {https://doi.org/10.1143/JPSJ.76.034701} {\bibfield  {journal} {\bibinfo
  {journal} {J. Phys. Soc. Jpn.}\ }\textbf {\bibinfo {volume} {76}},\ \bibinfo
  {pages} {034701} (\bibinfo {year} {2007})}\BibitemShut {NoStop}%
\bibitem [{\citenamefont {Maruyama}\ \emph {et~al.}(2012)\citenamefont
  {Maruyama}, \citenamefont {Sigrist},\ and\ \citenamefont
  {Yanase}}]{Maruyama2012JPSJ}%
  \BibitemOpen
  \bibfield  {author} {\bibinfo {author} {\bibfnamefont {D.}~\bibnamefont
  {Maruyama}}, \bibinfo {author} {\bibfnamefont {M.}~\bibnamefont {Sigrist}},\
  and\ \bibinfo {author} {\bibfnamefont {Y.}~\bibnamefont {Yanase}},\
  }\bibfield  {title} {\bibinfo {title} {Locally non-centrosymmetric
  superconductivity in multilayer systems},\ }\href@noop {} {\bibfield
  {journal} {\bibinfo  {journal} {J. Phys. Soc. Jpn}\ }\textbf {\bibinfo
  {volume} {81}},\ \bibinfo {pages} {034702} (\bibinfo {year}
  {2012})}\BibitemShut {NoStop}%
\bibitem [{\citenamefont {Hashimoto}\ \emph {et~al.}(2013)\citenamefont
  {Hashimoto}, \citenamefont {Yada}, \citenamefont {Yamakage}, \citenamefont
  {Sato},\ and\ \citenamefont {Tanaka}}]{HashimotoJPSJ2013}%
  \BibitemOpen
  \bibfield  {author} {\bibinfo {author} {\bibfnamefont {T.}~\bibnamefont
  {Hashimoto}}, \bibinfo {author} {\bibfnamefont {K.}~\bibnamefont {Yada}},
  \bibinfo {author} {\bibfnamefont {A.}~\bibnamefont {Yamakage}}, \bibinfo
  {author} {\bibfnamefont {M.}~\bibnamefont {Sato}},\ and\ \bibinfo {author}
  {\bibfnamefont {Y.}~\bibnamefont {Tanaka}},\ }\bibfield  {title} {\bibinfo
  {title} {Bulk electronic state of superconducting topological insulator},\
  }\href {https://doi.org/10.7566/JPSJ.82.044704} {\bibfield  {journal}
  {\bibinfo  {journal} {J. Phys. Soc. Jpn.}\ }\textbf {\bibinfo {volume}
  {82}},\ \bibinfo {pages} {044704} (\bibinfo {year} {2013})}\BibitemShut
  {NoStop}%
\bibitem [{\citenamefont {Kashiwaya}\ \emph {et~al.}(1999)\citenamefont
  {Kashiwaya}, \citenamefont {Tanaka}, \citenamefont {Yoshida},\ and\
  \citenamefont {Beasley}}]{KTYB99}%
  \BibitemOpen
  \bibfield  {author} {\bibinfo {author} {\bibfnamefont {S.}~\bibnamefont
  {Kashiwaya}}, \bibinfo {author} {\bibfnamefont {Y.}~\bibnamefont {Tanaka}},
  \bibinfo {author} {\bibfnamefont {N.}~\bibnamefont {Yoshida}},\ and\ \bibinfo
  {author} {\bibfnamefont {M.~R.}\ \bibnamefont {Beasley}},\ }\bibfield
  {title} {\bibinfo {title} {Spin current in
  ferromagnet-insulator-superconductor junctions},\ }\href@noop {} {\bibfield
  {journal} {\bibinfo  {journal} {Phys. Rev. B}\ }\textbf {\bibinfo {volume}
  {60}},\ \bibinfo {pages} {3572} (\bibinfo {year} {1999})}\BibitemShut
  {NoStop}%
\bibitem [{\citenamefont {Hirai}\ \emph {et~al.}(2003)\citenamefont {Hirai},
  \citenamefont {Tanaka}, \citenamefont {Yoshida}, \citenamefont {Asano},
  \citenamefont {Inoue},\ and\ \citenamefont {Kashiwaya}}]{Hirai03}%
  \BibitemOpen
  \bibfield  {author} {\bibinfo {author} {\bibfnamefont {T.}~\bibnamefont
  {Hirai}}, \bibinfo {author} {\bibfnamefont {Y.}~\bibnamefont {Tanaka}},
  \bibinfo {author} {\bibfnamefont {N.}~\bibnamefont {Yoshida}}, \bibinfo
  {author} {\bibfnamefont {Y.}~\bibnamefont {Asano}}, \bibinfo {author}
  {\bibfnamefont {J.}~\bibnamefont {Inoue}},\ and\ \bibinfo {author}
  {\bibfnamefont {S.}~\bibnamefont {Kashiwaya}},\ }\bibfield  {title} {\bibinfo
  {title} {Temperature dependence of spin-polarized transport in
  ferromagnet/unconventional superconductor junctions},\ }\href@noop {}
  {\bibfield  {journal} {\bibinfo  {journal} {Phys. Rev. B}\ }\textbf {\bibinfo
  {volume} {67}},\ \bibinfo {pages} {174501} (\bibinfo {year}
  {2003})}\BibitemShut {NoStop}%
\bibitem [{\citenamefont {Tanaka}\ and\ \citenamefont
  {Kashiwaya}(1995)}]{TK95}%
  \BibitemOpen
  \bibfield  {author} {\bibinfo {author} {\bibfnamefont {Y.}~\bibnamefont
  {Tanaka}}\ and\ \bibinfo {author} {\bibfnamefont {S.}~\bibnamefont
  {Kashiwaya}},\ }\bibfield  {title} {\bibinfo {title} {Theory of tunneling
  spectroscopy of $\mathit{d}$-wave superconductors},\ }\href@noop {}
  {\bibfield  {journal} {\bibinfo  {journal} {Phys. Rev. Lett.}\ }\textbf
  {\bibinfo {volume} {74}},\ \bibinfo {pages} {3451} (\bibinfo {year}
  {1995})}\BibitemShut {NoStop}%
\bibitem [{\citenamefont {Kashiwaya}\ and\ \citenamefont
  {Tanaka}(2000)}]{KashiwayaTanaka2000RepProgPhys}%
  \BibitemOpen
  \bibfield  {author} {\bibinfo {author} {\bibfnamefont {S.}~\bibnamefont
  {Kashiwaya}}\ and\ \bibinfo {author} {\bibfnamefont {Y.}~\bibnamefont
  {Tanaka}},\ }\bibfield  {title} {\bibinfo {title} {Tunnelling effects on
  surface bound states in unconventional superconductors},\ }\href
  {http://stacks.iop.org/0034-4885/63/i=10/a=202} {\bibfield  {journal}
  {\bibinfo  {journal} {Rep. Prog. Phys.}\ }\textbf {\bibinfo {volume} {63}},\
  \bibinfo {pages} {1641} (\bibinfo {year} {2000})}\BibitemShut {NoStop}%
\end{thebibliography}%

\end{document}